\PassOptionsToPackage{dvipsnames}{xcolor}
\documentclass[10pt,sigconf,nonacm,screen]{acmart}

\usepackage[all]{nowidow}
\usepackage{xcolor}
\usepackage{subcaption}
\usepackage{hyperref}
\usepackage{acronym}
\usepackage{tikz}
\usepackage{lipsum} % optional, for testing layout
\usepackage{tablefootnote}
\usepackage[capitalise,noabbrev]{cleveref}
\crefformat{section}{\S#2#1#3} % see manual of cleveref, section 8.2.1
\crefformat{subsection}{\S#2#1#3}
\crefformat{subsubsection}{\S#2#1#3}
\crefrangeformat{section}{\S#3#1#4 to~\S#5#2#6}
\crefrangeformat{subsection}{\S#3#1#4 to~\S#5#2#6}
\crefrangeformat{subsubsection}{\S#3#1#4 to~\S#5#2#6}

\usepackage{todonotes}
\definecolor{darkblue}{rgb}{0.0, 0.0, 0.55}
\begin{document}

% use a separate author block for each author, ACM wants that
% leave out the "\department" if you need space

% \author{Diana Baumann}
% \affiliation{%
%     \institution{TU Berlin}
%     % \department{Scalable Software Systems Research Group}
%     \city{Berlin}
%     \country{Germany}}
% \email{diba@3s.tu-berlin.de}
% \orcid{0009-0006-3000-6691}

% \author{Valentin Carl}
% \affiliation{%
%     \institution{TU Berlin}
%     % \department{Scalable Software Systems Research Group}
%     \city{Berlin}
%     \country{Germany}}
% \email{vc@3s.tu-berlin.de}
% \orcid{0009-0000-5991-9255}

% \author{Nils Japke}
% \affiliation{%
%     \institution{TU Berlin}
%     % \department{Scalable Software Systems Research Group}
%     \city{Berlin}
%     \country{Germany}}
% \email{nj@3s.tu-berlin.de}
% \orcid{0000-0002-2412-4513}

% \author{Niklas Kowallik}
% \affiliation{%
%     \institution{TU Berlin}
%     % \department{Scalable Software Systems Research Group}
%     \city{Berlin}
%     \country{Germany}}
% \email{nk@3s.tu-berlin.de}
% \orcid{0009-0006-9839-1864}

% \author{Mohammadreza Malekabbasi}
% \affiliation{%
%     \institution{TU Berlin}
%     % \department{Scalable Software Systems Research Group}
%     \city{Berlin}
%     \country{Germany}}
% \email{mm@3s.tu-berlin.de}

% \author{Tobias Pfandzelter}
% \affiliation{%
%     \institution{TU Berlin}
%     % \department{Scalable Software Systems Research Group}
%     \city{Berlin}
%     \country{Germany}}
% \email{tp@3s.tu-berlin.de}
% \orcid{0000-0002-7868-8613}

\author{Tim C. Rese}
\affiliation{%
    \institution{TU Berlin}
    % \department{Scalable Software Systems Research Group}
    \city{Berlin}
    \country{Germany}}
\email{tr@3s.tu-berlin.de}
\orcid{0009-0008-0185-8339}

% \author{Trever Schirmer}
% \affiliation{%
%     \institution{TU Berlin}
%     % \department{Scalable Software Systems Research Group}
%     \city{Berlin}
%     \country{Germany}}
% \email{ts@3s.tu-berlin.de}
% \orcid{0000-0001-9277-3032}

% \author{Minghe Wang}
% \affiliation{%
%     \institution{TU Berlin}
%     % \department{Scalable Software Systems Research Group}
%     \city{Berlin}
%     \country{Germany}}
% \email{mw@3s.tu-berlin.de}
% \orcid{0009-0001-3780-5828}

\author{Alexandra Kapp}
\affiliation{%
    \institution{TU Berlin}
    % \department{Scalable Software Systems Research Group}
    \city{Berlin}
    \country{Germany}}
\email{ak@3s.tu-berlin.de}
\orcid{0000-0002-8348-8958}

\author{David Bermbach}
\affiliation{%
    \institution{TU Berlin}
    % \department{Scalable Software Systems Research Group}
    \city{Berlin}
    \country{Germany}}
\email{db@3s.tu-berlin.de}
\orcid{0000-0002-7524-3256}

% ACM allows short titles for running heads
% but its only necessary if your title is too long or you need to add artificial newlines to the title on the front page that you don't want to appear on the subsequent pages
% \title[A Shorter Paper Template Title]{Paper Template: A Lightweight Paper Template}
\title{Evaluating the Impact Of Spatial Features Of Mobility Data and Index Choice On Database Performance}


% can ignore this if you have "nonacm" set
% once you publish, ACM will help you add stuff here
% set noacm class option

\copyrightyear{2025}
\acmYear{2025}
% \setcopyright{acmcopyright}\acmConference[Conference '24]{29th International Conference}{December 26--31, 2024}{Pyongyang, People's Republic of Korea}
% \acmBooktitle{29th International Conference (Conference '24), December 26--31, 2024, Pyongyang, People's Republic of Korea}

\begin{abstract}
    The growing number of moving Internet-of-Things (IoT) devices has led to a surge in moving object data, powering applications such as traffic routing, hotspot detection, or weather forecasting.
    When managing such data, spatial database systems offer various index options and data formats, e.g., point-based or trajectory-based.
		Likewise, dataset characteristics such as geographic overlap and skew can vary significantly.
		All three significantly affect database performance.
		While this has been studied in existing papers, none of them explore the effects and trade-offs resulting from a combination of all three aspects.		
    
		In this paper, we evaluate the performance impact of index choice, data format, and dataset characteristics on a popular spatial database system, PostGIS.
    We focus on two aspects of dataset characteristics, the degree of overlap and the degree of skew, and propose novel approximation methods to determine these features.
    We design a benchmark that compares a variety of spatial indexing strategies and data formats, while also considering the impact of dataset characteristics on database performance.
    We include a variety of real-world and synthetic datasets, write operations, and read queries to cover a broad range of scenarios that might occur during application runtime.

    Our results offer practical guidance for developers looking to optimize spatial storage and querying, while also providing insights into dataset characteristics and their impact on database performance.
\end{abstract}

\maketitle
\begin{tikzpicture}[remember picture,overlay]
    \node[anchor=south west, xshift=1.8cm, yshift=0.5cm] at (current page.south west) {
      \begin{minipage}[t]{0.45\textwidth}
      \footnotesize
      © 2025 IEEE. Personal use of this material is permitted. Permission from
      IEEE must be obtained for all other uses, in any current or future media,
      including reprinting/republishing this material for advertising or promotional
      purposes, creating new collective works, for resale or redistribution to servers
      or lists, or reuse of any copyrighted component of this work in other works.\\
      DOI: 10.1109/IC2E65552.2025.00007
      \end{minipage}
    };
    \end{tikzpicture}
    
\section{Introduction}
Spatial data has become integral to modern applications due to the ever-growing amount of Internet-of-Things (IoT) devices.
These devices produce geospatial information used in services such as real-time traffic-aware routing, incident hotspot detection, and weather updates and predictions, with various other use cases existing~\cite{liebig2017dynamic,karakaya2020simra,karakay2022sumo,mendelsohn2007climate}.
Building such applications requires informed decisions regarding the technology stack, especially regarding data storage factors. 

Spatial indexes enhance query performance in spatial databases by enabling efficient querying of spatial data~\cite{nguyen2009indexing, aref2001sp,zhu2007efficient}.
These structures provide a way to quickly locate spatially co-located data points and vectors, while also enabling filtering of irrelevant data points.
Some systems offer a single default index, while others, such as PostGIS, support multiple types, each with unique trade-offs between query speed, index creation time, and maintenance costs.

Moving object data is an important subset of spatial data, showing the movement of multiple objects over time.
Such data can be stored in different formats, such as discrete points or trajectories~\cite{zheng2015trajectory,wang2021survey,spaccapietra2008conceptual}.
While purpose-built systems for trajectories exist, general-purpose solutions such as PostGIS remain widely used due to their flexibility and ecosystem compatibility.

Choosing an optimal index and format for both spatial data and moving object data is non-trivial, as not all indexes are equally effective for all datasets and query types~\cite{nam2004comparative}.
This is important to consider, as the data format can impact the possible queries, the granularity of the query response, and the performance of the database.
Performance may also vary depending on the characteristics of the data, such as spatial distribution and amount of data overlap, which can be difficult to determine in advance.

This paper investigates how spatial data characteristics, data format, and index choice impact database performance using PostGIS as a prototype database platform for our evaluation.
We construct an application-driven benchmark using both synthetic and real-world spatial datasets with varying data distributions and degrees of overlap.
We provide novel approximation methods to determine the degree of skew and overlap, which scale for large datasets by introducing a Monte Carlo method which calculates the degrees within 3\% of their respective values using less than 1\% of the required time it takes to accurately calculate them. 
Our benchmark includes both read and write evaluations to fully regard the impact of dataset properties, index choice, and data format on database performance.
Based on our results, we provide guidance for developers seeking to make informed storage decisions tailored to their use case.

Our key contributions are as follows:
\begin{itemize}
    \item We develop novel metrics and tunable approximation methods to assess overlap and distribution properties of trajectory-based datasets (\cref{sec:approach}).
    \item We design a benchmark for comparing data formats, data characteristics,  and index types using both real-world and synthetic datasets in PostGIS, analyzing their impact on read and write performance (\cref{sec:experiment_design}).
    \item We provide practical recommendations which index type and data format to choose depending on one's dataset characteristics (\cref{sec:summary}).
\end{itemize}

% place table at the bottom of the page

\section{Background and Related Work}
\label{sec:background}
\begin{table*}[!ht]
    \centering
    \caption{Each of these indexing strategies has its own advantages and disadvantages, which can be used to determine the best index for a specific use case. The related work mentioned in the table is not exhaustive, but provides a good overview.}
    \begin{tabular}{|p{0.15\linewidth}|p{0.15\linewidth}|p{0.15\linewidth}|p{0.15\linewidth}|p{0.15\linewidth}|p{0.15\linewidth}|}
        \hline
        Index Type & R-Tree & BRIN & GiST & SP-GiST & Space-Filling Curves \\
        \hline
        Summary & Partitions points and spatial data in vector format based on their minimum bounding rectangle across a balanced tree structure & Relies on data being sorted by the queried attribute to optimally function due to dividing a table into block ranges & Uses a generalized R-Tree approach to function for multiple data types & Relies on an underlying QuadTree; Designed for space partitioned data (not necessarily spatial data) & Maps multiple dimensions of the dataset (do not have to be spatial/temporal) to a singular continuous curve \\
        \hline
        Commonly used for & Spatial data & Any data that can be sorted by the queried attribute & Various use cases & Spatially partitioned data & Multidimensional data \\
        \hline
        Used/Evaluated/ Adapted in & ~\cite{zhu2007efficient,balasubramanian2012state,guttman1984r} & ~\cite{wang2023slbrin,wu2017apply}  & ~\cite{hellerstein1995generalized,simion2013price,schoemans2024multi} & ~\cite{aref2001sp,schoemans2024multi} & ~\cite{lawder2000application,lawder2000using,wang2005space} \\
        \hline
    \end{tabular}
    
    \label{tab:indeces}
\end{table*}
In this section, we provide a background on moving object data, relevant storage and indexing strategies.
Lastly, we highlight key related work that guides our paper. 

\subsection*{Moving Object Data}
Moving object data is a category of spatial data that captures the position of objects over time enables mobility-based applications. such as traffic prediction, route planning, and fleet management~\cite{meng2011moving,gidofalvi2010using,guting2005moving}.
Typically, this data originates from GPS sensors, producing a sequence of point observations.
A trajectory represents a continuous path constructed by linking these points, for example, tracing a bicyclist's commute ~\cite{luca2021survey}.

\begin{figure*}
    \centering
    \includegraphics[width=0.8\textwidth]{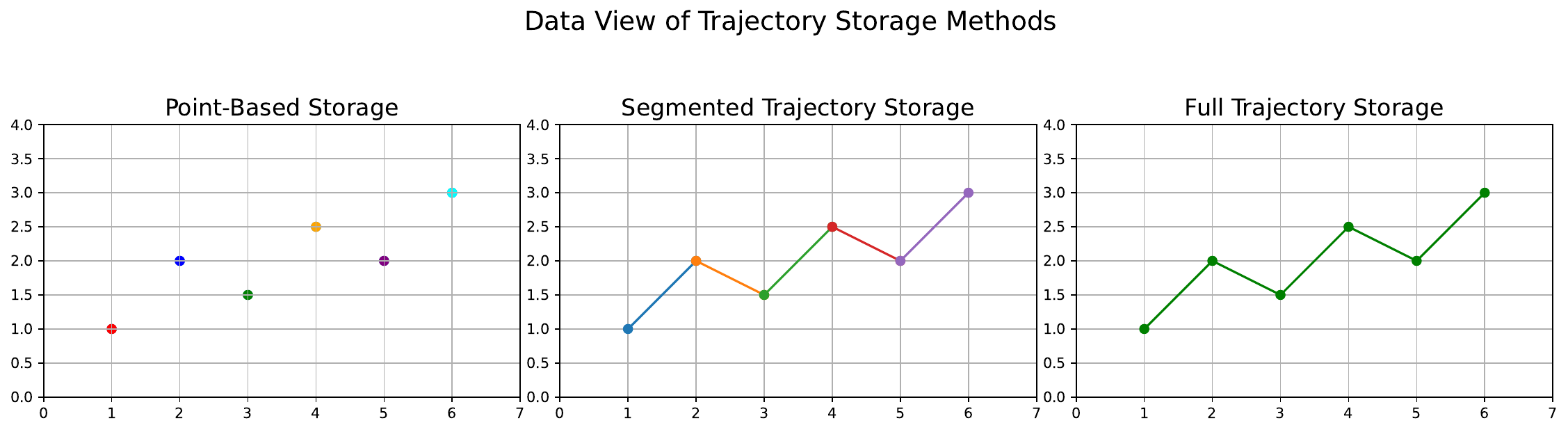}
    \caption{Moving Object Data can be stored in various formats, such as simply storing the point data. One can also store segments of the trajectory separately (each color represents a separate entry in the database), or store the entire trajectory as one object.}
    \label{fig:storage}
\end{figure*}

Trajectories can be stored in different formats, such as storing individual points, segments of a trajectory (each with its own entry), or storing the full trajectory as a single object. 
Each format offers trade-offs: Storing trajectory segments allows for more fine-grained analysis, while whole-trajectory storage simplifies representation and storage at the cost of query flexibility.
\cref{fig:storage} illustrates how these formats differ from one another.

Some databases provide interpolation features to support trajectory-based queries while storing data in a point-based format, with MobilityDB being a notable example~\cite{zimanyi2020mobilitydb}.
Here, different interpolation strategies can be applied, with linear interpolation being the most common.

\subsection*{Spatial Indexing Strategies}
Spatial Indexes exist to support efficient querying of spatial data.
These include R-Trees, QuadTrees, Generalized Search Trees (GiST), Block Range Indexes (BRIN), space-filling curves, and many more. Each of these have trade-offs in terms of data structure, dimensionality, and query performance.
Many of these indexes rely on bounding boxes such as the Minimum Bounding Rectangle (MBR) to quickly filter irrelevant data points.
A summary of common index types is shown in \cref{tab:indeces}, however, many more exist to also include other attributes such as time~\cite{hadjieleftheriou2002efficient,tian2022survey}.
Systems exist that implement custom indexing and storage techniques for trajectory data ~\cite{li2020trajmesa,cudre2010trajstore,bakli2019hadooptrajectory}.

\subsection*{Factors Affecting Performance}
Several features of trajectory data affect database performance, which includes the storage format, spatial distribution of data (data skew), and the degree of spatial overlap (intersections).
Storage format influences how efficiently queries can be processed.
For example, segmenting trajectories may enable better indexing but increase complexity.
Data skew refers to uneven distribution of data over space, such as traffic clustering in urban areas.
This may lead to an imbalance in indexing structures and performance degradation.
Intersections refer to overlapping spatial objects, and while related, they are distinct from skew.
High skew does not necessarily imply high intersection, as we show in \cref{sec:approach}.
For instance, delivery vehicles often operate in dense urban zones (high skew) without frequent overlap due to route optimization.

Studies have addressed both phenomena.
Chen et al. proposed specialized representations and intersection algorithms for 3D spatial data~\cite{chen2009data}.
Others explored detecting skew in systems such as SpatialHadoop~\cite{belussi2018detecting,zhang2010spatial}.

\subsection*{The Role of Representative Datasets}
Benchmarking indexing performance requires realistic, diverse datasets.
Using only one dataset, as is often done, fails to account for performance variations due to data characteristics. 

For example, Zhang et al. demonstrate how data skew can be exploited to optimize queries~\cite{zhang2020exploiting}.
Publicly available datasets like the Piraeus AIS maritime dataset provide ample real-world data (over 240 million records) for this purpose~\cite{tritsarolis2022piraeus}.

\subsection*{Related Work}
A wide range of research has investigated spatial indexes and benchmarks.
Nguyen et al. demonstrated the benefit of using spatial indexes with trajectories in PostGIS, but only evaluated GiST and ignored other strategies, and also only considered a single data format ~\cite{nguyen2009indexing}.
Additionally, the work focuses on a road network, which is not representative of moving object data in general.

Chen et al. benchmarked several indexes outside of a database, with synthetic car trajectory data~\cite{chen2008benchmark}.
However, they did not include real-world datasets or assess data formats.

Xu et al. proposed GMOBench, a benchmark for multi-modal trajectory data, but again relied on synthetic datasets and did not explore storage alternatives~\cite{xu2015gmobench}. 
This approach however did consider the possible impact of storing 3 or 4-dimensional trajectory data. 

BerlinMOD introduced a mobility benchmark focusing on spatiotemporal queries and data generation~\cite{duntgen2009berlinmod}.
Its contribution lies in query categorization, not in evaluating index or storage format impact.

TrajStore proposed a specialized trajectory storage format and indexing mechanism~\cite{cudre2010trajstore}.
While valuable, its architecture differs from widely-used systems like PostGIS.

\section{Approach}
\label{sec:approach}
Previous research has shown that certain indexes are better suited for overlapping and non-overlapping data, and that data distribution can also impact the performance of a database depending on the index~\cite{cutt2008improving,rabl2013variations}.
Developers therefore should be interested in their dataset characteristics to make an informed decision on which index to use.

In this section, we describe our benchmarking approach, while highlighting dataset features that we consider when selecting representative datasets for our evaluation.
We provide novel approximation methods to quantify overlap and distribution in a dataset, which can be used as a guideline for developers to choose the appropriate index for their data.
\subsection*{Assessing the Impact of Data Skew and Overlap}
Real datasets, while sharing common characteristics, can differ greatly in terms of data skew and overlap. 
Data such as urban cycling data is often more evenly distributed across the city, however, when including all data from a country, a large skew may be present.

\subsubsection*{Intersections and Overlaps}
\begin{figure}
    \centering
    \includegraphics[width=1.0\linewidth]{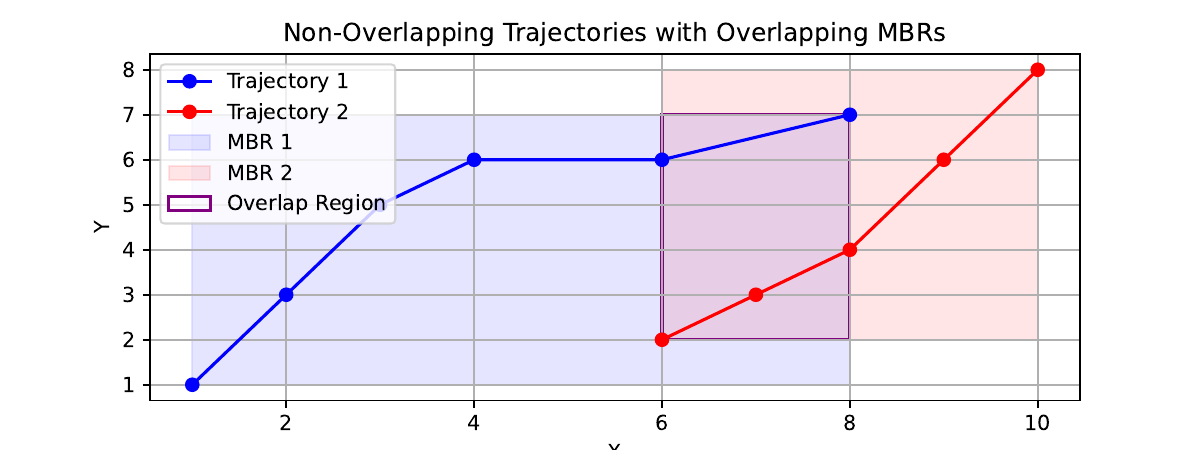}
    \caption{Trajectory 1 and 2 have overlapping minimum bounding rectangles, but do not intersect.}
    \label{fig:mbrexample}
\end{figure}
Intersections refer to two vectors or trajectories intersecting at one or more points, while overlaps refer to these trajectories occupying the same space for a larger amount of distance besides a single point. 
A simple example would be a 4-way street crossing: Two bicycles going in perpendicular directions have intersecting paths, while two going in the same direction overlap.
In spatial databases that use MBR-based indexing strategies however, these two terms come together due to the way indexes are structured.
As mentioned in the previous section, MBRs help quickly filter irrelevant data when running queries and allow for faster query responses.
When regarding two trajectories for a possible intersection, the system first checks these MBR structures for an overlap, leading to the interchangeable use of the two words in this context. 
\cref{fig:mbrexample} shows how non-intersecting data can have overlapping MBRs.
\subsubsection*{Determining the Global Overlap Coefficient}
\begin{figure}
    \centering
    \includegraphics[width=1.0\linewidth]{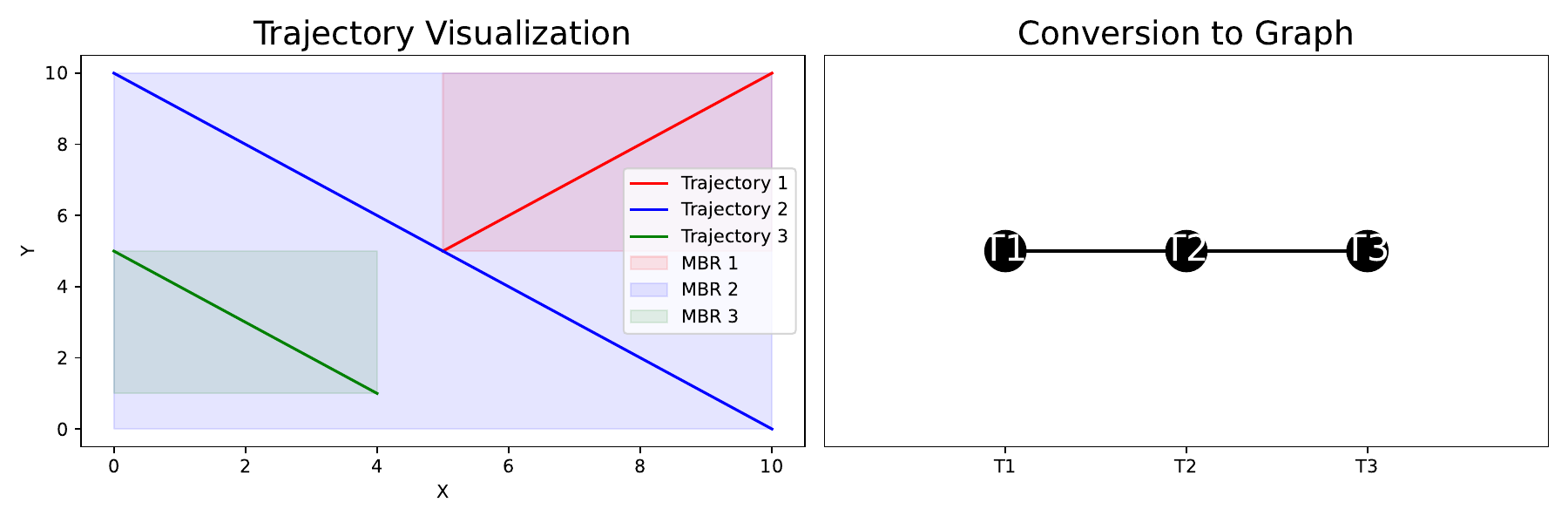}
    \caption{Traj. 1's MBR is overlapping with Traj. 2 and 3, while Traj. 2 and 3 are not overlapping. In graph form, each trajectory is a node and possesses an edge to overlapping trajectories. The GOC of this dataset would then be 2/3.}
    \label{fig:graphexample}
\end{figure}

The amount of overlap in a dataset is crucial for the choice of the most suitable index, as some indexes perform better when data is space partitioned ~\cite{aref2001sp}.
Given a trajectory \textit{t} and another trajectory \textit{t'}, we can determine if they overlap in matters of bounding-box-based-indexing by checking if their MBRs overlap.
Using this, we can turn our trajectory data into a graph to use an existing graph metric to determine the amount of overlap in our dataset.
We can represent each trajectory in our dataset as a node in a graph. 
If the MBR of two trajectories overlap, we create an edge between those nodes.
An example of how the trajectory data is converted to a graph can be seen in \cref{fig:graphexample}.
A graph representation of our data enables us to apply graph density measures to quantify the trajectory overlap. The graph density sets the number of total edges in relation to the number of all \textit{possible} edges.
Given a set of nodes $V$ and edges $E$, the density of a graph can be calculated as follows:
\[
D = \frac{2|E|}{|V|(|V|-1)}
\]
Going back to our way of converting trajectories to overlaps, density in our case reflects the amount of overlapping trajectories in relation to the total amount of possible overlaps, with its value ranging from 0 to 1.
A high density (i.e closer to 1) indicates that a high degree of trajectories overlap with one another, while a low value (closer to 0) indicates the opposite. 
The general graph approach is not without its problems, especially with highly overlapping data:
Generating a graph data structure from large trajctory datasets, containing millions or billions of instances, is computationally not feasible. 
We thus propose an approximation to allow developers a faster estimation of their dataset characteristics.
Given a dataset of \textit{m} trajectories:
\begin{itemize}
    \item We take $n$ randomly selected trajectories from our dataset. 
    \item We apply our approach to the selected trajectories and calculate the density of the graph.
    \item We repeat this process $p$ times and take the median of all approximated densities. 
\end{itemize}
This allows for a time complexity of $O(n^2\cdot p)$ instead of $O(m^2)$ for all graph sizes, where we can choose $n$ and $p$ based on the desired accuracy of our approximation.
Logically, the higher we set $n$ and $p$, the more accurate our approximation will be, at the cost of longer computation time.
In the rest of the paper, we refer to this metric as the global overlap coefficient (GOC) of a dataset. 

Calculating the exact GOC of a dataset, while possible, is often not feasible in a reasonable time frame. 
\cref{tab:approximation_goc} shows the results and calculation time of our GOC approximation method, highlighting the noticeable increase in calculation time with increasing $n$ and $p$ values. 
The highest approximation that we include ($n$ = 100, $p$ = 10,000) took over 2 hours to calculate, while the exact value took noticeably longer.
We recommend to not accurately calculate dataset GOC due to the time requirement, considering the likely sufficient approximation results. 
\subsubsection*{Adapting Average Nearest Neighbor to Trajectories}
The graph density, while suitable for evaluating overlap, does not cover the skew of a dataset.
Data in some scenarios can be highly clustered without overlapping, which our density coefficient would not be able to detect.
We therefore need an alternative approach to evaluate the distribution of our dataset. 
For point patterns, a common way to evaluate data distribution is using the average nearest neighbor (ANN) approach ~\cite{clark1954distance,thompson2022ancient}.
Here we determine the average distance of each point to its nearest neighbor. 
The result is compared to the expected value of a dataset with a uniform distribution. 
The observed distance can be calculated as follows:
\[
D_O = \frac{\sum_{i=1}^{n} d_i}{n}
\]
where $d_i$ is the distance of point $i$ to its nearest neighbor, and $n$ is the total amount of points in the dataset.
The expected distance in a uniformly distributed dataset with $n$ points can be calculated as:
\[
D_E = \frac{0.5}{\sqrt{n/A}}
\]
where $A$ is the area of the bounding box of the dataset.
The ANN is simply the ratio of the two values: 
\[
ANN = \frac{D_O}{D_E}
\]
A value of $<1$ indicates a clustered distribution, while a value of $>1$ indicates a distributed dataset. 
The larger the value, the more distributed the dataset is.
When applying this to trajectories, the idea does not hold up well: 
The distance between trajectories is the closest possible distance of the two, which is not indicative of their actual distribution.
Two trajectories may intersect at one point, but be very far apart otherwise and still have a distance of 0.
We can however still use the ANN idea if we convert our trajectories into multiple points. 
In this paper, we rely on the update frequency of the trajectory to convert trajectories, meaning that every time a sensor updates its position, we create a point in our dataset.
\begin{figure}
    \centering
    \includegraphics[width=1.0\linewidth]{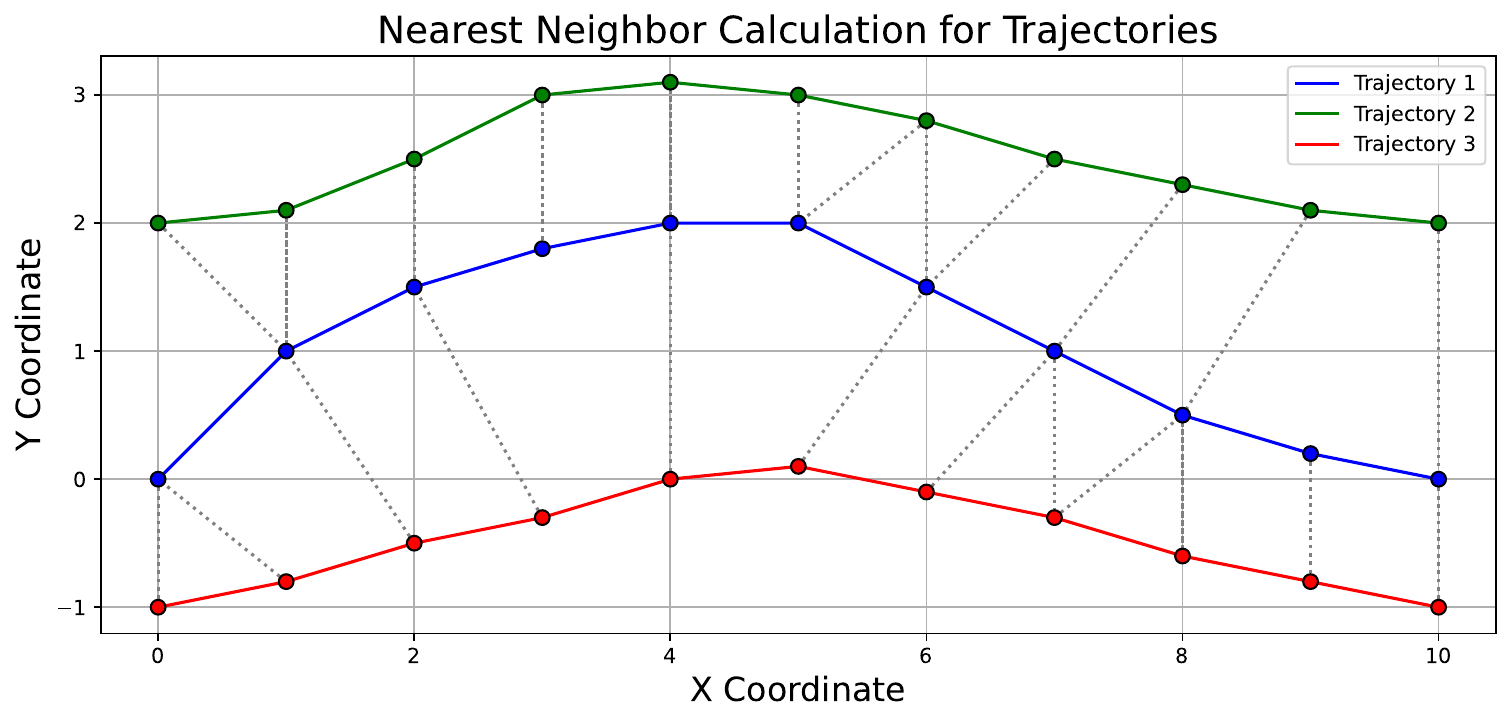}
    \caption{These simplified trajectories show how we can apply ANN to trajectories. With a small number of trajectories, exactly calculating this value is still realistic. Including a large amount of trajectories necessitates an approximation to finish the calculation in a reasonable time. Our ANN approximation excludes points from the own trajectory. }
    \label{fig:nearestneighbor}
\end{figure}
\begin{table*}[htb]
    \centering
    \caption{We use the aviation dataset also included in the evaluation to evaluate the accuracy of our approximation methods. Our approximations show that even with a small $n$ and $p$, we can achieve a near perfect approximation of the GOC and ANN values in a fraction of the required calculation time.}
    \label{tab:approximation}
    \begin{minipage}[t]{0.48\textwidth}
        \centering
        \caption*{(a) GOC}
        \begin{tabular}{|c|c|c|c|}
            \hline
            $n$ & $p$ & GOC & Calculation Time (s)\\
            \hline
            10 & 100 & 0.3759 & 4.25\\
            10 & 500 & 0.4083 & 20.64\\
            10 & 1,000 & 0.4017 & 50.22 \\
            10 & 10,000 & 0.4025 & 625.85 \\
            100 & 100 & 0.3985 & 51.70 \\
            100 & 500 & 0.3995 & 233.20 \\
            100 & 1,000 & 0.4005 & 548.87 \\
            100 & 10,000 & 0.4012 & 7525.20 ($\sim$2h) \\
            1 & 215,908 & 0.4012 & 96293.23 ($\sim$27h) \\
            \hline
        \end{tabular}
        \label{tab:approximation_goc}
    \end{minipage}%
    \hfill
    \begin{minipage}[t]{0.48\textwidth}
        \centering
        \caption*{(b) ANN}
        \begin{tabular}{|c|c|c|c|}
            \hline
            $n$ & $p$ & ANN & Calculation Time (s)\\
            \hline
            10 & 100 & 0.7002 & 2.65 \\
            10 & 500 & 0.6982 & 4.55 \\
            10 & 1,000 & 0.6943 & 11.19 \\
            10 & 10,000 & 0.6952 & 100.95 \\
            100 & 100 & 0.6908 & 9.1 \\
            100 & 500 & 0.6960 & 45.78 \\
            100 & 1,000 & 0.6954 & 92.3\\
            100 & 10,000 & 0.6942 & 918.96 \\
            1 & $\sim$30M & 0.6942 & 28621.15 ($\sim$8h) \\
            \hline
        \end{tabular}
        \label{tab:approximation_ann}
    \end{minipage}
\end{table*}

\cref{fig:nearestneighbor} shows how such a nearest neighbor approach would look like in a trajectory dataset, while immediately highlighting an issue:
Given a large amount of trajectories, which in turn is transformed into an even larger amount of points, the ANN approach is too computationally expensive to be used in a reasonable time frame.
We can again approximate the value by relying on a sampling strategy. 
Given a dataset containing $m$ trajectories, each with $k$ points, we use the following approach:
\begin{itemize}
    \item We take a random point from a random trajectory. 
    \item We calculate the nearest neighbor of this point, and store the distance.
    \item We repeat this process $n$ times, sum up the stored distances, and divide by $n$ to get the average distance.
    \item We scale this value by multiplying it with $(m\cdot k)/n$ to get an approximation of $D_O$ and calculate $D_E$ using the same formula as before (using the original amount of points in the dataset).
    \item We again repeat this process $p$ times and take the median of all results.
\end{itemize}
This allows us to reduce the time complexity to $O(n^2\cdot p)$, where we can choose $n$ and $p$ based on the desired accuracy of our approximation instead $O((m\cdot k)^2)$ for all dataset sizes.
We exclude all points from the own trajectory, as including these would not provide any valuable information regarding the dataset when evaluating trajectories.
We evaluate the accuracy of our approximations of GOC and ANN by comparing the approximated values to the exact results, using different $n$ and $p$ values to provide an overview of the accuracy of our methods.
\cref{tab:approximation_ann} shows the results of our approximation method.
Calculating the exact ANN value for a dataset containing $\sim$ 30 million points took over 8 hours, while our approximation method was able to achieve a near perfect approximation in a fraction of the time.
Assuming even larger datasets, we consider an approximation to be necessary to provide a reasonable time frame for developers to evaluate their dataset characteristics.

\subsubsection*{Covering all Bases with Representative Datasets}
We include real-world datasets from a variety of use cases, such as cycling data, aviation data, and ship trajectory (AIS) data.
However, when we want to evaluate the impact of skew and overlap on database performance, a broader combination of datasets is required. 
Various approaches to dataset generation are possible and have been implemented in a variety of papers ~\cite{duntgen2009berlinmod,kong2023mobility}.
By implementing the following dataset generation strategies, we can cover a broad range of use cases and data distributions:
\begin{itemize}
    \item Randomly generating trajectories within a bounding box.
    \item Evenly distributing trajectories within a bounding box, where we lay a raster over the bounding box and place trajectory starting points at the center of each cell.
    \item Enable a hotspot-based approach of trajectories, where trajectories mostly exist within hotspots and form clusters.
    \item Provide the same hotspot-approach with overlaps to other hotspots (as could be found in car traffic). 
\end{itemize}

We run our benchmark against all mentioned datasets, and evaluate the impact of dataset features on database performance using them. 
All included datasets can be found in \cref{tab:datasets}, where important data features are highlighted.
Data sets are stored in two different formats:
Line-based data can still differ in its storage format, as we could store the entire trajectory as a single object, or store each segment of the trajectory separately. 
These two approaches have trade-offs, with the segmented approach allowing for more fine-grained querying and analysis, while the trip as a whole reduces the entire trip to a single row in the database, which could lead to a change in performance.
Therefore, both formats are included in our benchmark.
A point-based approach reduces the amount of possible queries without prior conversion to a trajectory (at least within PostGIS and using real-world datasets), which led us to omit this format. 

\begin{figure*}[!ht]
    \centering
    \begin{subfigure}{\textwidth}
      \centering
      \includegraphics[width=\textwidth]{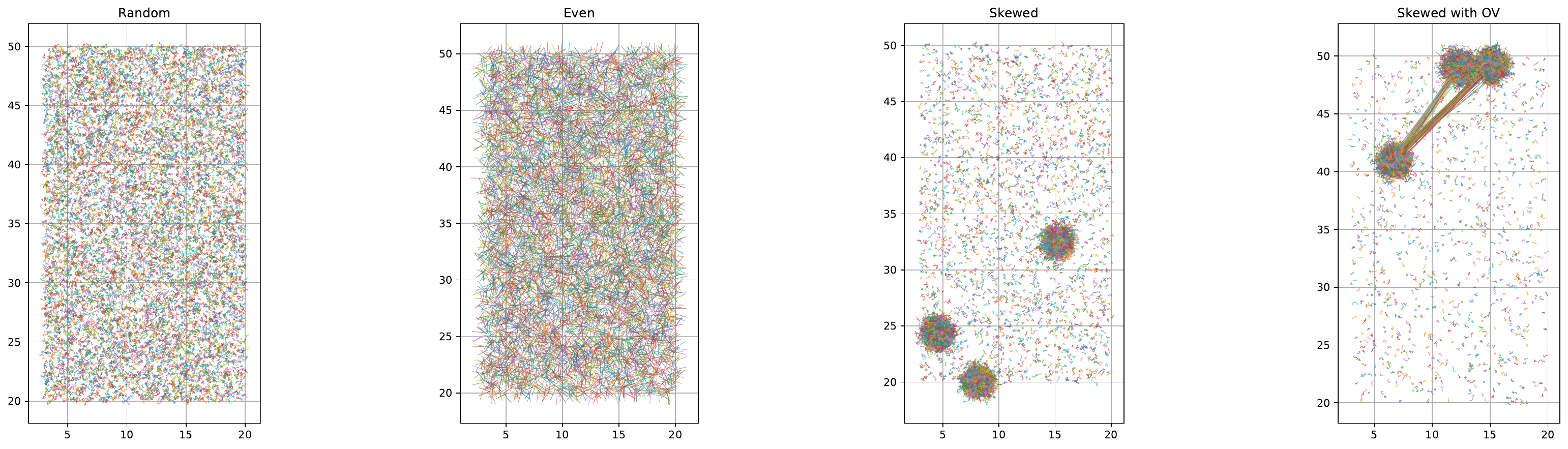}
      \label{fig:sub1}
    \end{subfigure}
    
    \vspace{0.01cm} % Adjust vertical spacing as needed
    
    \begin{subfigure}{\textwidth}
      \centering
      \includegraphics[width=\textwidth]{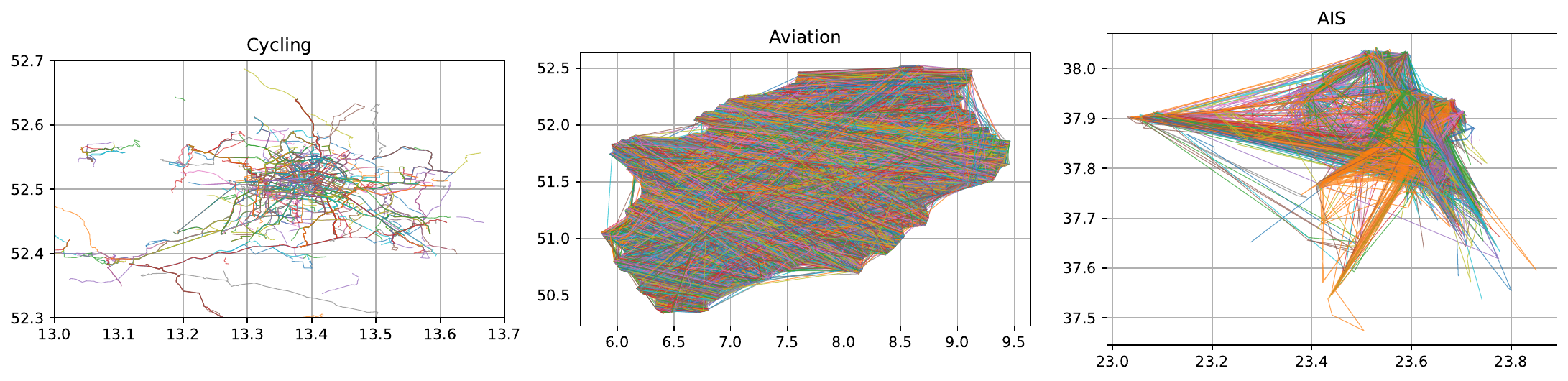}
      \label{fig:sub2}
    \end{subfigure}
    \caption{We include 7 different datasets in our evaluation, with 4 synthetic and 3 real-world datasets. The real-world datasets are from the SimRa project\protect\footnotemark[1], the Deutsche Flugsicherung\protect\footnotemark[2], and the Piraeus AIS dataset\protect\footnotemark[3].}
    \label{fig:datasets_visual}
\end{figure*}
\begin{table*}[!ht]
    \centering
    \caption{We include synthetic and real-world datasets to cover a variety of data patterns that might occur in various use cases. We calculate GOC and ANN using our approximation method where appropriate, doing 10 iterations of 10000 samples each. Each dataset consists of 30,000,000 segments, which are distributed across the listed number of trajectories. The name in parentheses is the name used in our evaluation.}
    \begin{tabular}{ |p{0.2\linewidth}|p{0.3\linewidth}|p{0.15\linewidth}|p{0.05\linewidth}|p{0.05\linewidth}| }
        \hline
        Dataset & Description  & \# Trajectories & GOC & ANN  \\
        \hline
        Random (Random) & Random trajectories & 3,000,000 & 0.0064 & 27.053 \\
        \hline
        Evenly Distributed (Even) & Evenly distributed trajectories & 3,000,000 & 0.0072 & 27.310 \\
        \hline
        Skewed (Skewed) & Clustered data, where a high percentage of trajectories are centered around hotspots & 3,000,000 & 0.048 & 16.477 \\
        \hline
        Skewed with Overlap (Skewed with OV) & Clustered data, where a high percentage of trajectories are centered around hotspots, and a degree of trajectories travel to other hotspots & 3,000,000 & 0.0553 & 12.963 \\
        \hline
        SimRa (Cycling)& Cycling trajectory data from Berlin and Potsdam & 4,387 & 0.212 & 5.052 \\
        \hline
        Deutsche Flugsicherung (Aviation)& Flight trajectory data from Northrhine-Westphalia & 215,908 & 0.4012 & 0.694 \\
        \hline
        Piraeus AIS Dataset (AIS)& AIS ship data from a research maritime dataset & 2,719 & 0.784 & 0.996 \\
        \hline
    \end{tabular}
    \label{tab:datasets}
\end{table*}
\section{Evaluation}
\label{sec:evaluation}
In this section, we evaluate database performance using both the synthetic and real-world datasets, and regard the impact of data format, index choice, and dataset characteristics in both read and write scenarios.
\subsection{Experiment Design}
\label{sec:experiment_design}
\begin{figure*}[!ht]
    \centering
    \begin{subfigure}{0.5\textwidth}
      \centering
      \includegraphics[width=0.85\linewidth]{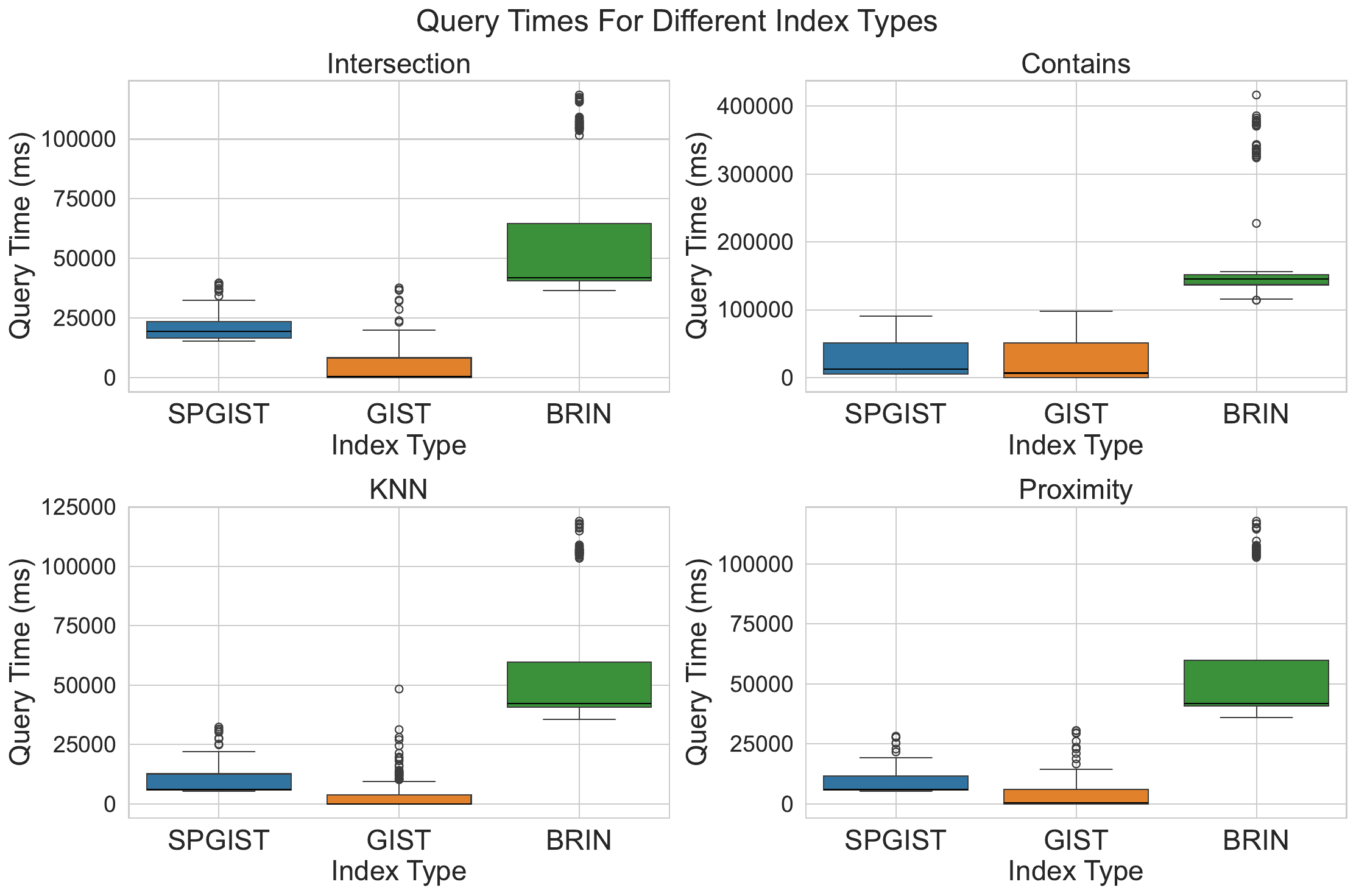}
      \caption{Segmented data}
      \label{fig:sub1}
    \end{subfigure}%
    \begin{subfigure}{0.5\textwidth}
      \centering
      \includegraphics[width=0.85\linewidth]{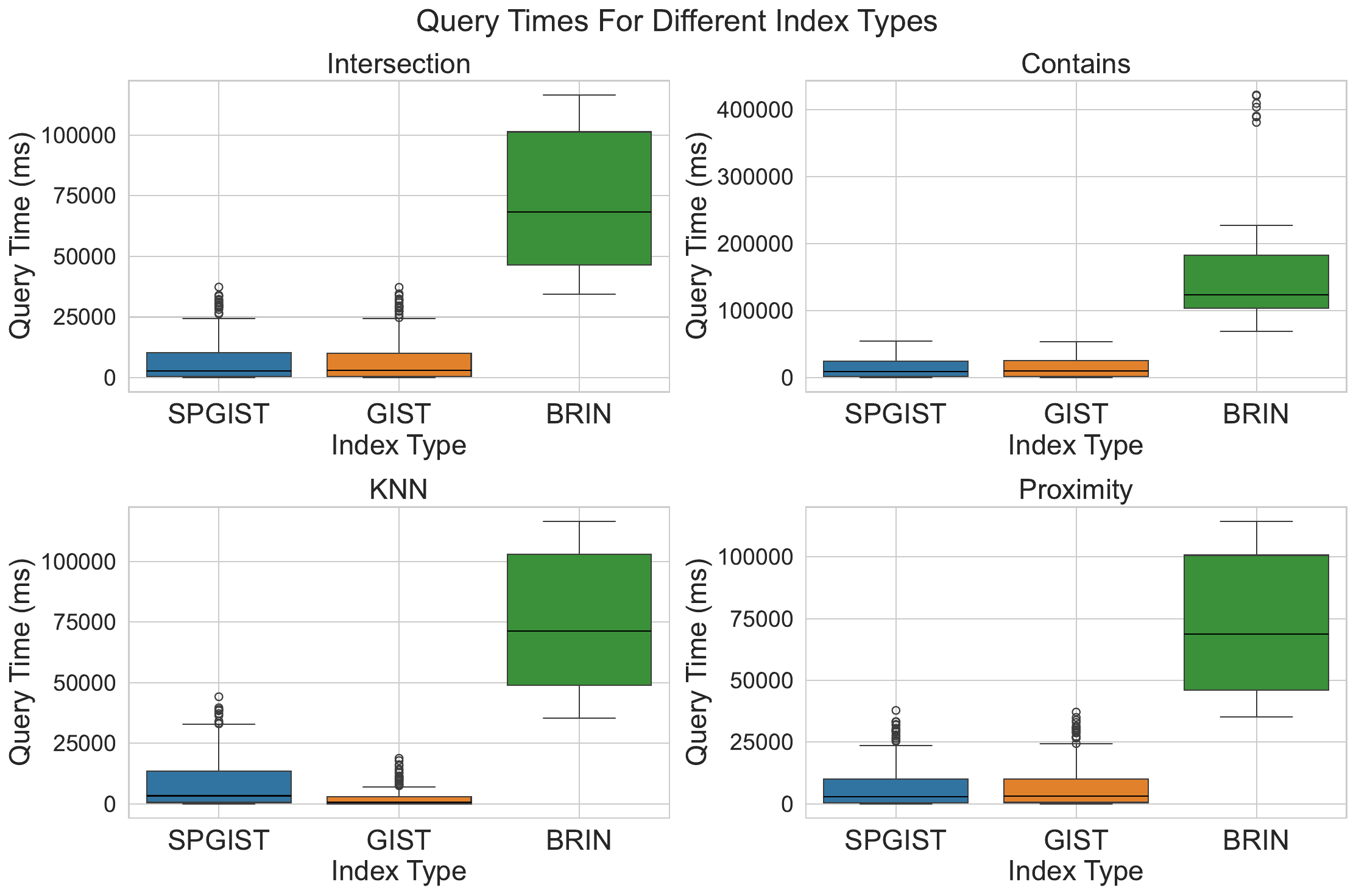}
      \caption{Non-segmented data}
      \label{fig:sub2}
    \end{subfigure}
    \caption{On average, GiST outperforms SP-GiST, especially in the segmented formats, however, there are scenarios where SP-GiST does provide an advantage. If designing a general purpose application able to handle a variety of queries, GiST is the better choice. BRIN performed poorly across all datasets and query types, and should likely not be used for trajectory data.}
    \label{fig:segment_traj_boxplot}
\end{figure*}
We include different read query types in our evaluation to cover a broad range of real-world applications.
\begin{itemize}
    \item How many other trajectories does a given trajectory overlap with (\textit{Intersection} query)? 
    \item What trajectories are partially/completely within a specified polygon (\textit{Contains} query)?
    \item What are my \text{K} nearest neighbors to a given trajectory (\textit{KNN} query)? 
    \item How many trajectories are within a specific distance of a given trajectory (\textit{Proximity} query)?
\end{itemize}

We additionally evaluate the three different write operations that can be performed (\textit{Insert}, \textit{Update}, and \textit{Delete}), as they cover basic application scenarios and index choice may impact performance here as well.
For our write operations, we insert either a single trajectory or 100 trajectories into the database.
Regarding the update and delete scenarios, we either adjust a single trajectory or 1\% of the dataset in a batch operation.

Each of these queries is run with 50 unique configurations. Using a \textit{Contains} and a single \textit{Insert} query as an example: 
During each of the configurations, the bounding box of the polygon in which trajectories are queried is unique and a randomly generated trajectory is inserted into the database within a specified polygon.
The system under test (SUT) is therefore subjected to 50 different benchmark configurations for each query type.
We take further considerations into account here, such as filtering polygons which return no results using rejection sampling, and ensuring that the bounding box is within the area of the dataset.
Each of these experiments is run against each dataset in every unique combination of index type and data format that we include in this paper. 
Each dataset is included in two formats, a segmented and non-segmented version, and three different index types are included in the evaluation (GiST, SP-GiST, and BRIN).
As SP-GiST is specifically designed for space partitioned data, our expectation  is that datasets with low overlap will benefit from this index type. 
The segmented version of the dataset is created by splitting the trajectory using the update frequency of the dataset.
\footnotetext[1]{\url{https://github.com/simra-project/dataset}}
\footnotetext[2]{The dataset is not open-source, but can be requested from the Deutsche Flugsicherung.}
\footnotetext[3]{\url{https://zenodo.org/records/6323416}}

To fairly compare the impact of spatial features and data format, our dataset size is fixed across all datasets.
We always include 30, 000, 000 segments of trajectories across each dataset to ensure that we can fairly evaluate the impact of overlap/distribution.
The key difference in datasets is how these are distributed across single entire trajectories, and how they are distributed across their respective bounding boxes.

\begin{figure*}[!ht]
    \centering
    \includegraphics[width=1.0\textwidth]{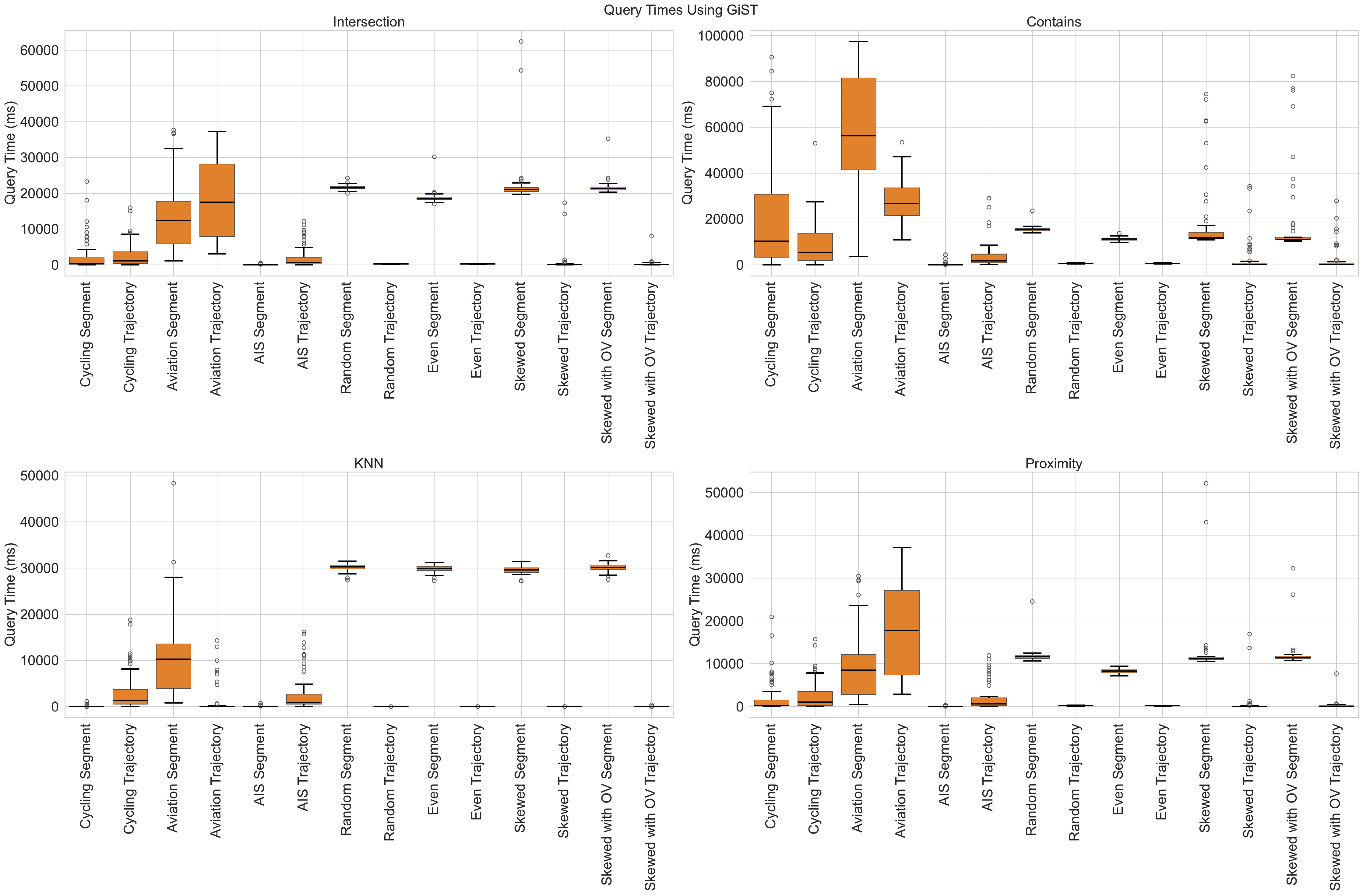}
    \caption{The GOC of a dataset here seems to be a good indicator whether the non-segmented format has any benefit at all, with our low GOC datasets showing performance benefits for the non-segmented format, while the ones with a higher GOC experience better performance with the segmented format. The \textit{Contains} query type is the only one where the non-segmented format performs better across almost all datasets.}  
    \label{fig:format}
\end{figure*}
\subsection{Experiment Setup}
All of our experiments were run on a 8-core Intel Xeon 4310 CPU with 32GB of RAM, with the database being run as a single instance.
The SUT uses PostGIS 3.5.0 and PostgreSQL 16.8, with the database being run on a single instance. 
During initialization, we deploy all datasets to the SUT, with index creation happening before the evaluation.

\subsection{Experiment Results}
We first evaluate the read performance of our datasets using a variety of queries, and afterwards run write experiments, where we run separate insert, update, and delete benchmark runs. 
Within each part, we evaluate the impact of data format, index choice, and dataset characteristics.

\subsubsection{Read Performance}
We ran three experiment repetitions, but due to the high amount of results, we will focus on one run.
The other runs were similar in their results.
\paragraph*{Impact of Spatial Index Choice}
Results from our evaluation show GiST and SP-GiST to be the best performing index types for our read queries, with GiST outperforming SP-GiST especially in the non-segmented format. 
This was the likely result, as the larger bounding boxes resulting from the non-segmented format do not allow for efficient space partitioned indexing of our data, which is the main advantage of SP-GiST.

\cref{fig:segment_traj_boxplot} shows the performance of the different index types across all datasets and query types, with the segmented data on the left and the non-segmented data on the right.
For trajectory data in PostGIS, BRIN performed poorly across all datasets and query types.
The trajectory format makes it inefficient to use BRIN, and while it may be beneficial for point data in some cases, our results show that one should likely not use BRIN when working with trajectory data.
\paragraph*{Impact of Data Format and Characteristics}

\cref{fig:format} highlights the performance difference for all query types for both formats using GiST, as it was the best performing index type overall. 
Our results show a split between the two formats, with all of our high GOC datasets performing better with the segmented format, while the low GOC datasets benefitted from the non-segmented format.

\begin{figure}[!ht]
    \centering
    \includegraphics[width=1.0\linewidth]{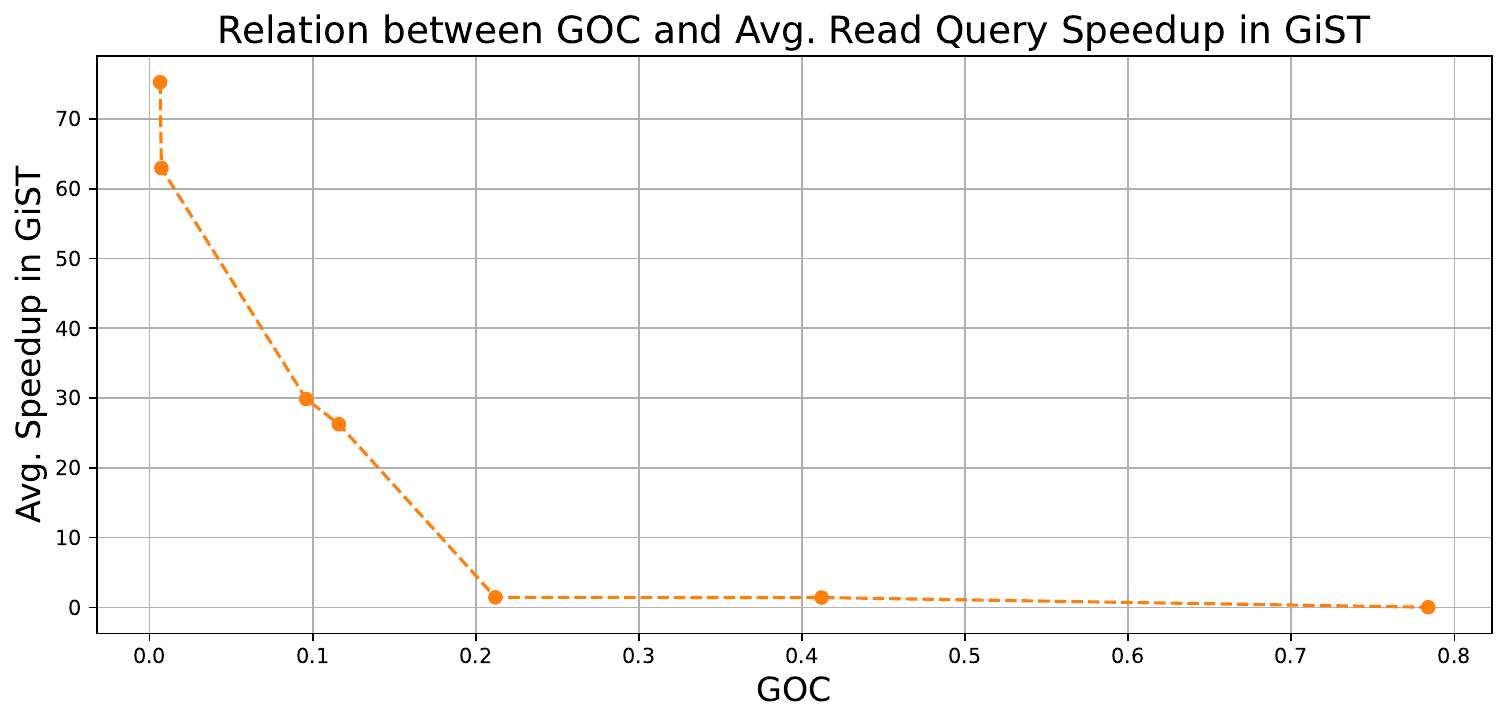}
    \caption{In our evaluation, datasets with a lower GOC benefitted heavily from using the non-segmented format when averaging across all query types. Higher GOC datasets received little to no speedup, with AIS data even performing better when using the segmented format. We were not able to determine a significant correlation however ($p=0.0564, \alpha=0.05$).}
    \label{fig:goc_GiST_relation}
\end{figure}

This lead us to further investigate the relation between GOC and the average speedup from relying on the non-segmented format when using GiST as an index type when averaging across all query types.
\cref{fig:goc_GiST_relation} shows the relation between GOC and average speedup across all query types when using GiST as an index type, with the dotted line indicating a possible interpolation between values. 
We find that the lower the GOC of a dataset, the better the performance of the non-segmented format, with the high GOC datasets performing better with the segmented format across most query types.
Further research here is required to determine the exact relation between GOC and performance, with one idea being to determine the elbow method for the GOC values to determine a threshold where the non-segmented format is beneficial.

While our ANN coefficient is able to provide insights into the distribution of the data, it does not seem to impact performance in our evaluation regarding the read performance overall.
Regardless, we believe that it is a valuable addition and could potentially provide insights into performance when relying on other databases or indexing strategies.
\begin{figure*}[!ht]
    \centering
    \includegraphics[width=1.0\textwidth]{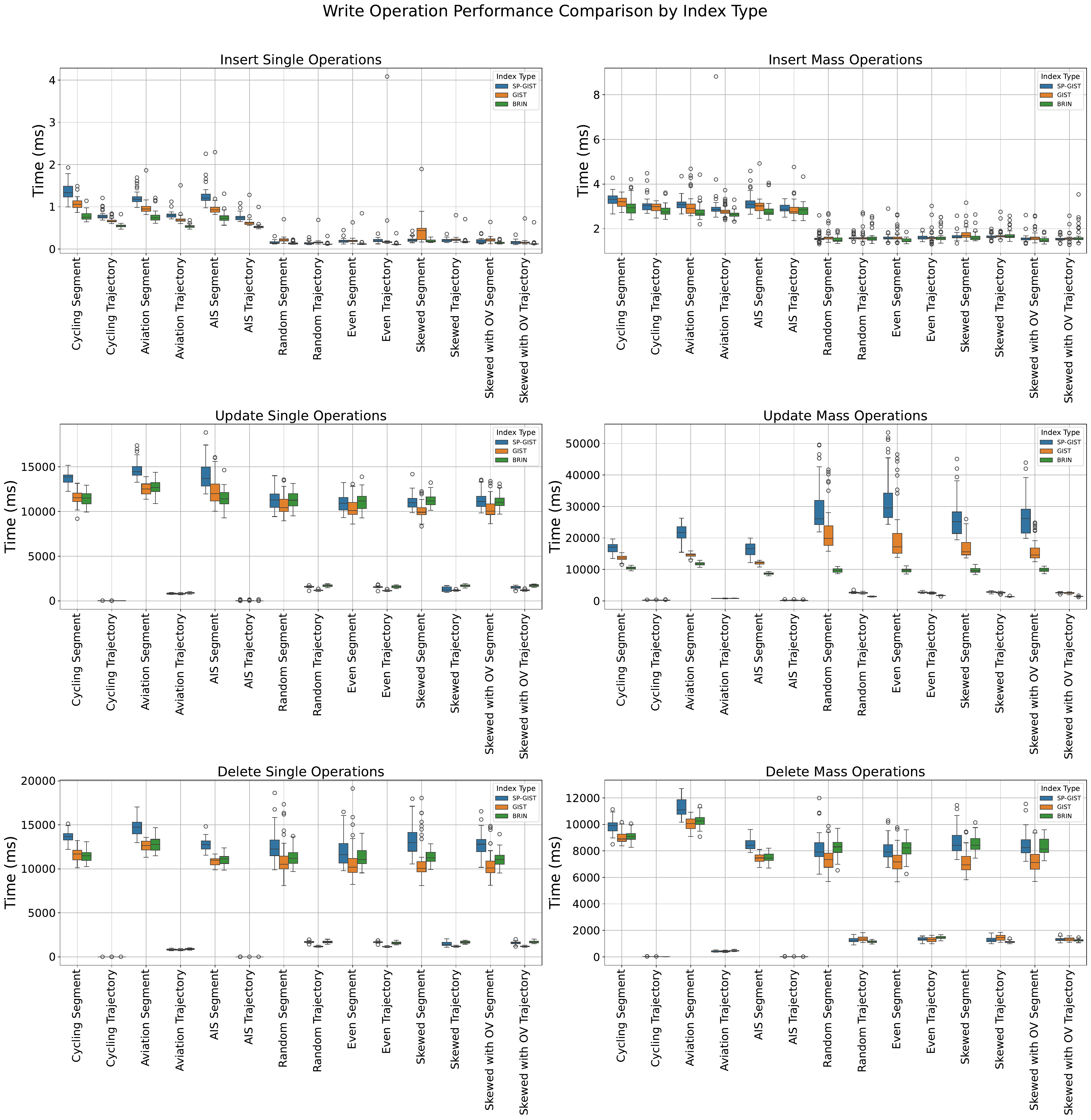}
    \caption{BRIN is usually the best performing index type for most write operations. SP-GiST was outperformed by GiST in nearly all of our cases, however, it sometimes provides a small advantage in datasets with low overlap. Of note is the noticeably better performance in our low GOC datasets for insert operations.}
    \label{fig:write_boxplot}
\end{figure*}
\subsubsection{Write Performance}
Our write experiments highlight how data format plays a large role in database performance across all three types of write operations, while also showing that BRIN can provide an advantage in some cases.
\paragraph*{Impact of Index Choice}
\cref{fig:write_boxplot} shows the performance of the different index types across all datasets and query types, with the singular operations on the left and the batch operations on the right.
Our results show that BRIN is the best performing index type for most write operations, however, the benefit is not as pronounced as it was negative for read operations.
As its benefits are inconsistently spread across write operations, we believe that it is not a good choice for trajectory data in PostGIS unless an application is heavily write focused.
SP-GiST was outperformed by GiST in nearly all of our cases, however, it sometimes provides a small advantage in datasets with low overlap. 
In single insert scenarios, SP-GiST was able to outperform GiST on average in all datasets, however, the performance difference was negligible.
\paragraph*{Impact of Data Format and Characteristics Impact}
When regarding the impact of data format on write performance, we found that the segmented format lagged behind the non-segmented trajectory in nearly all of our comparisons. 
This is as expected, due to the fact that each operation on a segmented trajectory requires the database to perform the operation on multiple rows, whether it be inserting a trajectory in a segmented format or deleting/updating an existing one.
The difference is not as pronounced in our insert operations, as we do not insert a percentage of the dataset, but a fixed number of trajectories. 
\subsection{Summary of Findings \& Recommendations for Developers}
\label{sec:summary}
\begin{figure*}[!ht]
    \centering
    \includegraphics[width=1.0\linewidth]{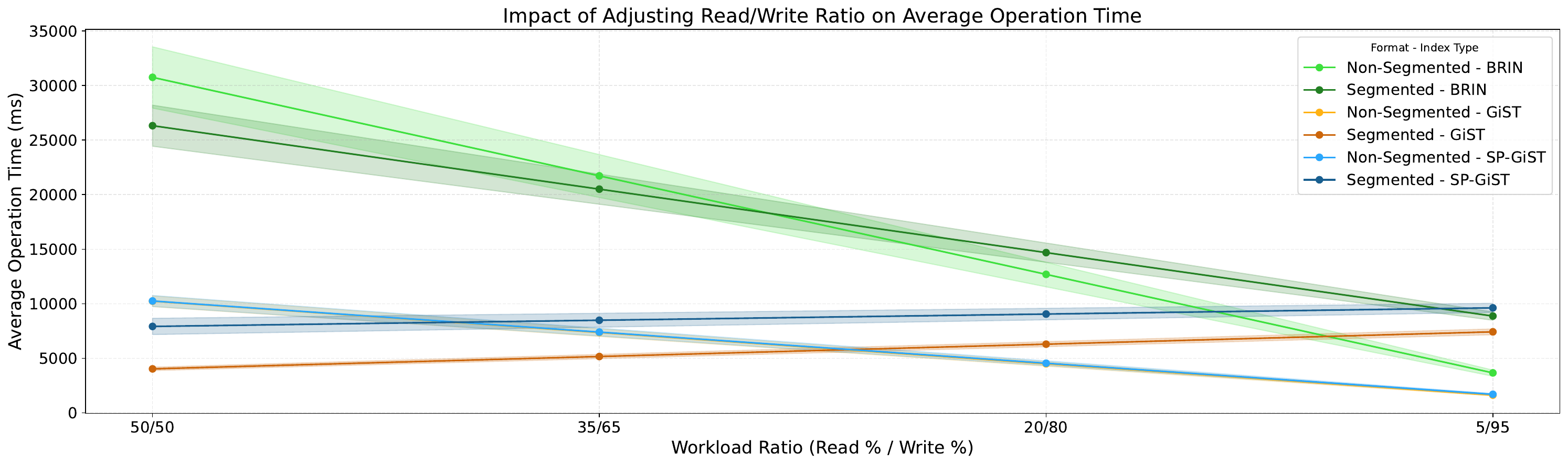}
    \caption{Unless the expected workload is extremely write-heavy, relying on BRIN is not recommended. The shaded area around the mean indicates the 95\% confidence interval, and the yellow line is nearly identical to the light blue one. The mean and interval are calculated across all datasets and query types.}
    \label{fig:avg_read_write_time}
\end{figure*}
When specifically regarding the impact of GOC and ANN on write performance, we focus on the segmented format, as the number of segments is fixed here across all datasets.
Insert operations, where a bounding box is used to insert new trajectories, show that the GOC may have an impact on performance, as our higher GOC datasets performed noticeably worse than the lower GOC datasets. 
However, these are not in ascending order, and we were not able to determine a correlation between GOC and performance.

While we still believe trajectory ANN to be an important metric, it did not impact performance in our evaluation. 
When relying on another indexing strategy and database, it may be beneficial to include ANN in the evaluation again to regard a possible correlation with performance.

Our results show that GiST remains the dominant choice when wanting to index spatial trajectory data in PostGIS, as it outperforms  or performs similarly to SP-GiST in nearly all cases, while outperforming it in nearly all scenarios with non-segmented data. 
When developers are deciding how to store their data, we provide the following recommendations based on our results:
When storing non-segmented data, GiST remains the optimal choice within the scope of our evaluation, as it outperforms SP-GiST in all cases. 

When implementing a segmented data format, both index types perform similarly, with SP-GiST performing slightly better in some cases. These cases are however, not tied to the degree of overlap in the data.

Our findings show that the higher the GOC of the data, the less of a benefit it is to store data in a non-segmented format. When using GiST, high GOC datasets benefit only slightly and in some cases even suffer from the non-segmented format. For those datasets, we recommend using the segmented format, as it provides better performance and a higher level of detail.
While adapting ANN to trajectories provides novel insights into a dataset, we did not find a correlation between ANN and performance. 
While other experiments may show a relationship here, our results indicate that data skew does not impact performance when using trajectory-based data in PostGIS.
ANN however, could be interesting to regard when evaluating other databases and/or indexing strategies present in such database systems. 

When running a write-heavy workload, the non-segmented format provides an advantage as it allows for faster writes across all datasets. 
While BRIN indexes are a good choice if one only considers write performance, its poor read performance likely makes it unsuitable for most applications. 
BRIN may show better read performance in scenarios where data is inserted  in a sorted order, i.e, with an always increasing timestamp, and then running time-based queries.
However, as such data is not trajectory-based, we did not include it in our evaluation.
% \begin{table*}[!ht]
%     \centering
%     \caption{The better write performance of BRIN likely will not balance out the benefits of the other two index types unless an application is extremely write-heavy.}
%     \begin{tabular}{|c|c|c|c|c|}
%         \hline
%         Format & Index Type & Average Read Time (ms) & Average Write Time (ms) & Sum (ms) \\
%         \hline
%         Non-Segmented & GiST & 8011.69 & 672.94 & 8684.63 \\
%         \hline
%         Non-Segmented & SP-GiST & 9485.22 & 751.78 & 10237 \\
%         \hline
%         Non-Segmented & BRIN & 90926.86 & 656.49 & 91583.35 \\
%         \hline
%         Segmented & GiST & 10089.66 & 7805.78 & 17895.44 \\
%         \hline
%         Segmented & SP-GiST & 17370.70 & 9813.46 & 27184.16 \\
%         \hline
%         Segmented & BRIN & 88549.92 & 6934.8 & 95484.72 \\
%         \hline
%     \end{tabular}
%     \label{tab:avg_read_write_time}
% \end{table*}

If expecting a mix of read and write queries, we recommend evaluating the GOC of the data to determine the optimal data format for performance, and relying on GiST as the index type.
We recommend to also keep the level of required detail in mind, and to regard if the non-segmented format is at all beneficial or usable for the intended application. 
Our expectation  was that SP-GiST would prove a better choice in scenarios where data is equally distributed across the observed area, as it was specifically designed for cases where data is spatially partitioned.
However, results show that at least for trajectory data in  PostGIS, this is not the case.
Regarding write performance, both index types perform similarly in most cases, with BRIN having an advantage in most scenarios.
As previously mentioned, the downside of BRIN is in its read performance, which likely does not balance out the performance benefits even in write-heavy workloads.

In \cref{fig:avg_read_write_time}, we show that if one's application is not extremely write-heavy, the performance benefits of BRIN do not outweigh the read performance benefits of GiST and SP-GiST.
Even in the best case scenario (5\%/95\% Read/Write Ratio) for BRIN that we evaluated, the mean operation time is still lower for GiST in both cases and for SP-GiST in the non-segmented format.

The GOC seems to be a good indicator whether the non-segmented format has any benefit at all, while the ANN coefficient did not seem to impact performance in our evaluation. 
We assume that other indexing methods and datasets outside of the evaluated ones may show a correlation between ANN and performance, specifically those which are designed for non-skewed data.

\section{Discussion \& Future Work}
\label{sec:discussion}
In this section, we discuss limitations of our evaluation and suggest future work to address these.

While MobilityDB could be considered a more suitable SUT due to its focus on mobility data, we chose PostGIS as it is more widely used, and our evaluation is limited to spatial queries. 
MobilityDB, despite its spatiotemporal capabilities, does not introduce novel spatial indexing strategies and offers no significant advantage in purely spatial scenarios without temporal aspects. 
Future work should consider extending the evaluation to spatiotemporal queries, in which case MobilityDB or similar databases would be more appropriate. 

We focus on GiST, SP-GiST, and BRIN indexes, as these are the primary spatial indexing methods available in PostGIS. 
While additional strategies such as space-filling curves or alternative indexing approaches could offer further insights, initial tests showed negligible benefit or poor performance compared to the selected methods. 
Future evaluations could explore these approaches in different database systems that offer a broader range of indexing options. 

Our benchmark includes diverse datasets and query types to ensure general applicability. 
Nevertheless, certain edge cases and access patterns may not be fully represented. 
While we equalize datasets by segment count to ensure fairness, this may unintentionally bias the results.
We plan on evaluating the impact of using other equalization strategies, such as trajectory length or amount of trajectories in future work.
We mitigate this by using multiple configurations and repeating experiments.
Future work could expand the dataset range and normalization strategies to further reduce bias. 

\section{Conclusion}
\label{sec:conclusion}
In this paper, we regarded the performance of a popular spatial database and how the choice of index, data format, and dataset characteristics impacted the performance of a database, as these factors have been shown to impact the performance of a database in previous work but have not been considered together in detail.

For this, we designed novel metrics and approximation methods to determine the degree of skew and overlap in a dataset, which can be used to determine key data characteristics.
We use these in combination with a self-designed benchmark to evaluate the performance of a popular spatial database, using both synthetic and real-world datasets.
Our findings showed that especially data format and index choice can have a strong impact both on read and write performance, with data characteristics possibly influencing a possible speedup to be gained when relying on a specific data format.

While we did not find a correlation for our ANN coefficient and performance, we believe that it provides a novel insight into trajectory distribution and may be useful for further research with other benchmark configurations.
Future work should focus on extending our benchmark to further databases and indexing strategies to regard if data characteristics impact the performance of other databases and indexing strategies, while also including further datasets.

% this won't appear if you have "anonymous" set
% select what you need!
% \begin{acks}
%     % 6G_NeXt
%     Funded by the \grantsponsor{BMBF}{Bundesministerium für Bildung und Forschung (BMBF, German Federal Ministry of Education and Research)}{https://www.bmbf.de/bmbf/en} -- \grantnum{BMBF}{16KISK183}.
%     % OptiFaaS
%     Funded by the \grantsponsor{DFG}{Deutsche Forschungsgemeinschaft (DFG, German Research Foundation)}{https://www.dfg.de/en/} -- \grantnum{DFG}{495343202}.
%     % CoDes, EMPIRIS, SPENCER
%     Partially funded by the \grantsponsor{BMBF}{Bundesministerium für Bildung und Forschung (BMBF, German Federal Ministry of Education and Research)}{https://www.bmbf.de/bmbf/en} in the scope of the Software Campus 3.0 (Technische Universit\"at Berlin) program -- \grantnum{BMBF}{01IS23068}.
% \end{acks}

% it is recommended to balance your columns. if this command does not work properly, remove it and instead add the "pbalance" option for the document class
\balance

\bibliographystyle{ACM-Reference-Format}
\bibliography{bibliography.bib}

%%% -*-BibTeX-*-
%%% Do NOT edit. File created by BibTeX with style
%%% ACM-Reference-Format-Journals [18-Jan-2012].

\begin{thebibliography}{44}

%%% ====================================================================
%%% NOTE TO THE USER: you can override these defaults by providing
%%% customized versions of any of these macros before the \bibliography
%%% command.  Each of them MUST provide its own final punctuation,
%%% except for \shownote{}, \showDOI{}, and \showURL{}.  The latter two
%%% do not use final punctuation, in order to avoid confusing it with
%%% the Web address.
%%%
%%% To suppress output of a particular field, define its macro to expand
%%% to an empty string, or better, \unskip, like this:
%%%
%%% \newcommand{\showDOI}[1]{\unskip}   % LaTeX syntax
%%%
%%% \def \showDOI #1{\unskip}           % plain TeX syntax
%%%
%%% ====================================================================

\ifx \showCODEN    \undefined \def \showCODEN     #1{\unskip}     \fi
\ifx \showDOI      \undefined \def \showDOI       #1{#1}\fi
\ifx \showISBNx    \undefined \def \showISBNx     #1{\unskip}     \fi
\ifx \showISBNxiii \undefined \def \showISBNxiii  #1{\unskip}     \fi
\ifx \showISSN     \undefined \def \showISSN      #1{\unskip}     \fi
\ifx \showLCCN     \undefined \def \showLCCN      #1{\unskip}     \fi
\ifx \shownote     \undefined \def \shownote      #1{#1}          \fi
\ifx \showarticletitle \undefined \def \showarticletitle #1{#1}   \fi
\ifx \showURL      \undefined \def \showURL       {\relax}        \fi
% The following commands are used for tagged output and should be
% invisible to TeX
\providecommand\bibfield[2]{#2}
\providecommand\bibinfo[2]{#2}
\providecommand\natexlab[1]{#1}
\providecommand\showeprint[2][]{arXiv:#2}

\bibitem[Aref and Ilyas(2001)]%
        {aref2001sp}
\bibfield{author}{\bibinfo{person}{Walid~G Aref} {and} \bibinfo{person}{Ihab~F Ilyas}.} \bibinfo{year}{2001}\natexlab{}.
\newblock \showarticletitle{Sp-gist: An extensible database index for supporting space partitioning trees}.
\newblock \bibinfo{journal}{\emph{Journal of Intelligent Information Systems}}  \bibinfo{volume}{17} (\bibinfo{year}{2001}), \bibinfo{pages}{215--240}.
\newblock


\bibitem[Bakli et~al\mbox{.}(2019)]%
        {bakli2019hadooptrajectory}
\bibfield{author}{\bibinfo{person}{Mohamed Bakli}, \bibinfo{person}{Mahmoud Sakr}, {and} \bibinfo{person}{Taysir Hassan~A Soliman}.} \bibinfo{year}{2019}\natexlab{}.
\newblock \showarticletitle{HadoopTrajectory: a Hadoop spatiotemporal data processing extension}.
\newblock \bibinfo{journal}{\emph{Journal of geographical systems}}  \bibinfo{volume}{21} (\bibinfo{year}{2019}), \bibinfo{pages}{211--235}.
\newblock


\bibitem[Balasubramanian and Sugumaran(2012)]%
        {balasubramanian2012state}
\bibfield{author}{\bibinfo{person}{Lakshmi Balasubramanian} {and} \bibinfo{person}{M Sugumaran}.} \bibinfo{year}{2012}\natexlab{}.
\newblock \showarticletitle{A state-of-art in R-tree variants for spatial indexing}.
\newblock \bibinfo{journal}{\emph{International Journal of Computer Applications}} \bibinfo{volume}{42}, \bibinfo{number}{20} (\bibinfo{year}{2012}), \bibinfo{pages}{35--41}.
\newblock


\bibitem[Belussi et~al\mbox{.}(2018)]%
        {belussi2018detecting}
\bibfield{author}{\bibinfo{person}{Alberto Belussi}, \bibinfo{person}{Sara Migliorini}, {and} \bibinfo{person}{Ahmed Eldawy}.} \bibinfo{year}{2018}\natexlab{}.
\newblock \showarticletitle{Detecting skewness of big spatial data in SpatialHadoop}. In \bibinfo{booktitle}{\emph{Proceedings of the 26th ACM SIGSPATIAL International Conference on Advances in Geographic Information Systems}}. \bibinfo{pages}{432--435}.
\newblock


\bibitem[Chen et~al\mbox{.}(2008)]%
        {chen2008benchmark}
\bibfield{author}{\bibinfo{person}{Su Chen}, \bibinfo{person}{Christian~S Jensen}, {and} \bibinfo{person}{Dan Lin}.} \bibinfo{year}{2008}\natexlab{}.
\newblock \showarticletitle{A benchmark for evaluating moving object indexes}.
\newblock \bibinfo{journal}{\emph{Proceedings of the VLDB Endowment}} \bibinfo{volume}{1}, \bibinfo{number}{2} (\bibinfo{year}{2008}), \bibinfo{pages}{1574--1585}.
\newblock


\bibitem[Chen and Schneider(2009)]%
        {chen2009data}
\bibfield{author}{\bibinfo{person}{Tao Chen} {and} \bibinfo{person}{Markus Schneider}.} \bibinfo{year}{2009}\natexlab{}.
\newblock \showarticletitle{Data structures and intersection algorithms for 3d spatial data types}. In \bibinfo{booktitle}{\emph{Proceedings of the 17th ACM SIGSPATIAL International Conference on Advances in Geographic Information Systems}}. \bibinfo{pages}{148--157}.
\newblock


\bibitem[Clark and Evans(1954)]%
        {clark1954distance}
\bibfield{author}{\bibinfo{person}{Philip~J Clark} {and} \bibinfo{person}{Francis~C Evans}.} \bibinfo{year}{1954}\natexlab{}.
\newblock \showarticletitle{Distance to nearest neighbor as a measure of spatial relationships in populations}.
\newblock \bibinfo{journal}{\emph{Ecology}} \bibinfo{volume}{35}, \bibinfo{number}{4} (\bibinfo{year}{1954}), \bibinfo{pages}{445--453}.
\newblock


\bibitem[Cudre-Mauroux et~al\mbox{.}(2010)]%
        {cudre2010trajstore}
\bibfield{author}{\bibinfo{person}{Philippe Cudre-Mauroux}, \bibinfo{person}{Eugene Wu}, {and} \bibinfo{person}{Samuel Madden}.} \bibinfo{year}{2010}\natexlab{}.
\newblock \showarticletitle{Trajstore: An adaptive storage system for very large trajectory data sets}. In \bibinfo{booktitle}{\emph{2010 IEEE 26th International Conference on Data Engineering (ICDE 2010)}}. IEEE, \bibinfo{pages}{109--120}.
\newblock


\bibitem[Cutt and Lawrence(2008)]%
        {cutt2008improving}
\bibfield{author}{\bibinfo{person}{Bryce Cutt} {and} \bibinfo{person}{Ramon Lawrence}.} \bibinfo{year}{2008}\natexlab{}.
\newblock \showarticletitle{Improving join performance for skewed databases}. In \bibinfo{booktitle}{\emph{2008 Canadian Conference on Electrical and Computer Engineering}}. IEEE, \bibinfo{pages}{000387--000392}.
\newblock


\bibitem[D{\"u}ntgen et~al\mbox{.}(2009)]%
        {duntgen2009berlinmod}
\bibfield{author}{\bibinfo{person}{Christian D{\"u}ntgen}, \bibinfo{person}{Thomas Behr}, {and} \bibinfo{person}{Ralf~Hartmut G{\"u}ting}.} \bibinfo{year}{2009}\natexlab{}.
\newblock \showarticletitle{BerlinMOD: a benchmark for moving object databases}.
\newblock \bibinfo{journal}{\emph{The VLDB Journal}}  \bibinfo{volume}{18} (\bibinfo{year}{2009}), \bibinfo{pages}{1335--1368}.
\newblock


\bibitem[Gidofalvi and Ehsan(2010)]%
        {gidofalvi2010using}
\bibfield{author}{\bibinfo{person}{Gy{\"o}z{\"o} Gidofalvi} {and} \bibinfo{person}{Saqib Ehsan}.} \bibinfo{year}{2010}\natexlab{}.
\newblock \showarticletitle{Using Trajectories of Moving Objects in Traffic Prediction and Management}. In \bibinfo{booktitle}{\emph{Workshop on Movement Pattern Analysis 2010, Zurich, Switzerland, September 14, 2010}}. \bibinfo{pages}{6}.
\newblock


\bibitem[G{\"u}ting and Schneider(2005)]%
        {guting2005moving}
\bibfield{author}{\bibinfo{person}{Ralf~Hartmut G{\"u}ting} {and} \bibinfo{person}{Markus Schneider}.} \bibinfo{year}{2005}\natexlab{}.
\newblock \bibinfo{booktitle}{\emph{Moving objects databases}}.
\newblock \bibinfo{publisher}{Academic Press}.
\newblock


\bibitem[Guttman(1984)]%
        {guttman1984r}
\bibfield{author}{\bibinfo{person}{Antonin Guttman}.} \bibinfo{year}{1984}\natexlab{}.
\newblock \showarticletitle{R-trees: A dynamic index structure for spatial searching}. In \bibinfo{booktitle}{\emph{Proceedings of the 1984 ACM SIGMOD international conference on Management of data}}. \bibinfo{pages}{47--57}.
\newblock


\bibitem[Hadjieleftheriou et~al\mbox{.}(2002)]%
        {hadjieleftheriou2002efficient}
\bibfield{author}{\bibinfo{person}{Marios Hadjieleftheriou}, \bibinfo{person}{George Kollios}, \bibinfo{person}{Vassilis~J Tsotras}, {and} \bibinfo{person}{Dimitrios Gunopulos}.} \bibinfo{year}{2002}\natexlab{}.
\newblock \showarticletitle{Efficient indexing of spatiotemporal objects}. In \bibinfo{booktitle}{\emph{International conference on extending database technology}}. Springer, \bibinfo{pages}{251--268}.
\newblock


\bibitem[Hellerstein et~al\mbox{.}(1995)]%
        {hellerstein1995generalized}
\bibfield{author}{\bibinfo{person}{Joseph~M Hellerstein}, \bibinfo{person}{Jeffrey~F Naughton}, {and} \bibinfo{person}{Avi Pfeffer}.} \bibinfo{year}{1995}\natexlab{}.
\newblock \bibinfo{booktitle}{\emph{Generalized search trees for database systems}}.
\newblock \bibinfo{publisher}{September}.
\newblock


\bibitem[Karakaya et~al\mbox{.}(2020)]%
        {karakaya2020simra}
\bibfield{author}{\bibinfo{person}{Ahmet-Serdar Karakaya}, \bibinfo{person}{Jonathan Hasenburg}, {and} \bibinfo{person}{David Bermbach}.} \bibinfo{year}{2020}\natexlab{}.
\newblock \showarticletitle{SimRa: Using crowdsourcing to identify near miss hotspots in bicycle traffic}.
\newblock \bibinfo{journal}{\emph{Pervasive and Mobile Computing}}  \bibinfo{volume}{67} (\bibinfo{year}{2020}), \bibinfo{pages}{101197}.
\newblock


\bibitem[Karakaya et~al\mbox{.}(2022)]%
        {karakay2022sumo}
\bibfield{author}{\bibinfo{person}{Ahmet-Serdar Karakaya}, \bibinfo{person}{Konstantin K{\"o}hler}, \bibinfo{person}{Julian Heinovski}, \bibinfo{person}{Falko Dressler}, {and} \bibinfo{person}{David Bermbach}.} \bibinfo{year}{2022}\natexlab{}.
\newblock \showarticletitle{A Realistic Cyclist Model for SUMO Based on the SimRa Dataset}. In \bibinfo{booktitle}{\emph{Proceedings of the 20th Mediterranean Communication and Computer Networking Conference}} (Pafos, Cyprus) \emph{(\bibinfo{series}{MedComNet 2022})}. \bibinfo{publisher}{IEEE}, \bibinfo{address}{New York, NY, USA}, \bibinfo{pages}{166--173}.
\newblock
\urldef\tempurl%
\url{https://doi.org/10.1109/MedComNet55087.2022.9810439}
\showDOI{\tempurl}


\bibitem[Kong et~al\mbox{.}(2023)]%
        {kong2023mobility}
\bibfield{author}{\bibinfo{person}{Xiangjie Kong}, \bibinfo{person}{Qiao Chen}, \bibinfo{person}{Mingliang Hou}, \bibinfo{person}{Hui Wang}, {and} \bibinfo{person}{Feng Xia}.} \bibinfo{year}{2023}\natexlab{}.
\newblock \showarticletitle{Mobility trajectory generation: a survey}.
\newblock \bibinfo{journal}{\emph{Artificial Intelligence Review}} \bibinfo{volume}{56}, \bibinfo{number}{Suppl 3} (\bibinfo{year}{2023}), \bibinfo{pages}{3057--3098}.
\newblock


\bibitem[Lawder(2000)]%
        {lawder2000application}
\bibfield{author}{\bibinfo{person}{Jonathan~K Lawder}.} \bibinfo{year}{2000}\natexlab{}.
\newblock \emph{\bibinfo{title}{The application of space-filling curves to the storage and retrieval of multi-dimensional data}}.
\newblock \bibinfo{thesistype}{Ph.\,D. Dissertation}. \bibinfo{school}{Birkbeck, University of London, UK}.
\newblock


\bibitem[Lawder and King(2000)]%
        {lawder2000using}
\bibfield{author}{\bibinfo{person}{Jonathan~K Lawder} {and} \bibinfo{person}{Peter~JH King}.} \bibinfo{year}{2000}\natexlab{}.
\newblock \showarticletitle{Using space-filling curves for multi-dimensional indexing}. In \bibinfo{booktitle}{\emph{British National Conference on Databases}}. Springer, \bibinfo{pages}{20--35}.
\newblock


\bibitem[Li et~al\mbox{.}(2020)]%
        {li2020trajmesa}
\bibfield{author}{\bibinfo{person}{Ruiyuan Li}, \bibinfo{person}{Huajun He}, \bibinfo{person}{Rubin Wang}, \bibinfo{person}{Sijie Ruan}, \bibinfo{person}{Yuan Sui}, \bibinfo{person}{Jie Bao}, {and} \bibinfo{person}{Yu Zheng}.} \bibinfo{year}{2020}\natexlab{}.
\newblock \showarticletitle{Trajmesa: A distributed nosql storage engine for big trajectory data}. In \bibinfo{booktitle}{\emph{2020 IEEE 36th International Conference on Data Engineering (ICDE)}}. IEEE, \bibinfo{pages}{2002--2005}.
\newblock


\bibitem[Liebig et~al\mbox{.}(2017)]%
        {liebig2017dynamic}
\bibfield{author}{\bibinfo{person}{Thomas Liebig}, \bibinfo{person}{Nico Piatkowski}, \bibinfo{person}{Christian Bockermann}, {and} \bibinfo{person}{Katharina Morik}.} \bibinfo{year}{2017}\natexlab{}.
\newblock \showarticletitle{Dynamic route planning with real-time traffic predictions}.
\newblock \bibinfo{journal}{\emph{Information Systems}}  \bibinfo{volume}{64} (\bibinfo{year}{2017}), \bibinfo{pages}{258--265}.
\newblock


\bibitem[Luca et~al\mbox{.}(2021)]%
        {luca2021survey}
\bibfield{author}{\bibinfo{person}{Massimiliano Luca}, \bibinfo{person}{Gianni Barlacchi}, \bibinfo{person}{Bruno Lepri}, {and} \bibinfo{person}{Luca Pappalardo}.} \bibinfo{year}{2021}\natexlab{}.
\newblock \showarticletitle{A survey on deep learning for human mobility}.
\newblock \bibinfo{journal}{\emph{ACM Computing Surveys (CSUR)}} \bibinfo{volume}{55}, \bibinfo{number}{1} (\bibinfo{year}{2021}), \bibinfo{pages}{1--44}.
\newblock


\bibitem[Mendelsohn et~al\mbox{.}(2007)]%
        {mendelsohn2007climate}
\bibfield{author}{\bibinfo{person}{Robert Mendelsohn}, \bibinfo{person}{Pradeep Kurukulasuriya}, \bibinfo{person}{Alan Basist}, \bibinfo{person}{Felix Kogan}, {and} \bibinfo{person}{Claude Williams}.} \bibinfo{year}{2007}\natexlab{}.
\newblock \showarticletitle{Climate analysis with satellite versus weather station data}.
\newblock \bibinfo{journal}{\emph{Climatic Change}} \bibinfo{volume}{81}, \bibinfo{number}{1} (\bibinfo{year}{2007}), \bibinfo{pages}{71--83}.
\newblock


\bibitem[Meng and Chen(2011)]%
        {meng2011moving}
\bibfield{author}{\bibinfo{person}{Xiaofeng Meng} {and} \bibinfo{person}{Jidong Chen}.} \bibinfo{year}{2011}\natexlab{}.
\newblock \bibinfo{booktitle}{\emph{Moving objects management}}.
\newblock \bibinfo{publisher}{Springer}.
\newblock


\bibitem[Nam and Sussman(2004)]%
        {nam2004comparative}
\bibfield{author}{\bibinfo{person}{Beomseok Nam} {and} \bibinfo{person}{Alan Sussman}.} \bibinfo{year}{2004}\natexlab{}.
\newblock \showarticletitle{A comparative study of spatial indexing techniques for multidimensional scientific datasets}. In \bibinfo{booktitle}{\emph{Proceedings. 16th International Conference on Scientific and Statistical Database Management, 2004.}} IEEE, \bibinfo{pages}{171--180}.
\newblock


\bibitem[Nguyen(2009)]%
        {nguyen2009indexing}
\bibfield{author}{\bibinfo{person}{Thanh~Thi Nguyen}.} \bibinfo{year}{2009}\natexlab{}.
\newblock \showarticletitle{Indexing PostGIS databases and spatial Query performance evaluations}.
\newblock \bibinfo{journal}{\emph{International Journal of Geoinformatics}} \bibinfo{volume}{5}, \bibinfo{number}{3} (\bibinfo{year}{2009}), \bibinfo{pages}{1}.
\newblock


\bibitem[Rabl et~al\mbox{.}(2013)]%
        {rabl2013variations}
\bibfield{author}{\bibinfo{person}{Tilmann Rabl}, \bibinfo{person}{Meikel Poess}, \bibinfo{person}{Hans-Arno Jacobsen}, \bibinfo{person}{Patrick O'Neil}, {and} \bibinfo{person}{Elizabeth O'Neil}.} \bibinfo{year}{2013}\natexlab{}.
\newblock \showarticletitle{Variations of the star schema benchmark to test the effects of data skew on query performance}. In \bibinfo{booktitle}{\emph{Proceedings of the 4th ACM/SPEC International Conference on Performance Engineering}}. \bibinfo{pages}{361--372}.
\newblock


\bibitem[Schoemans et~al\mbox{.}(2024)]%
        {schoemans2024multi}
\bibfield{author}{\bibinfo{person}{Maxime Schoemans}, \bibinfo{person}{Walid~G Aref}, \bibinfo{person}{Esteban Zim{\'a}nyi}, {and} \bibinfo{person}{Mahmoud Sakr}.} \bibinfo{year}{2024}\natexlab{}.
\newblock \showarticletitle{Multi-Entry Generalized Search Trees for Indexing Trajectories}. In \bibinfo{booktitle}{\emph{Proceedings of the 32nd ACM International Conference on Advances in Geographic Information Systems}}. \bibinfo{pages}{421--431}.
\newblock


\bibitem[Simion et~al\mbox{.}(2013)]%
        {simion2013price}
\bibfield{author}{\bibinfo{person}{Bogdan Simion}, \bibinfo{person}{Daniel~N Ilha}, \bibinfo{person}{Angela~Demke Brown}, {and} \bibinfo{person}{Ryan Johnson}.} \bibinfo{year}{2013}\natexlab{}.
\newblock \showarticletitle{The price of generality in spatial indexing}. In \bibinfo{booktitle}{\emph{Proceedings of the 2nd ACM SIGSPATIAL International Workshop on Analytics for Big Geospatial Data}}. \bibinfo{pages}{8--12}.
\newblock


\bibitem[Spaccapietra et~al\mbox{.}(2008)]%
        {spaccapietra2008conceptual}
\bibfield{author}{\bibinfo{person}{Stefano Spaccapietra}, \bibinfo{person}{Christine Parent}, \bibinfo{person}{Maria~Luisa Damiani}, \bibinfo{person}{Jose~Antonio de Macedo}, \bibinfo{person}{Fabio Porto}, {and} \bibinfo{person}{Christelle Vangenot}.} \bibinfo{year}{2008}\natexlab{}.
\newblock \showarticletitle{A conceptual view on trajectories}.
\newblock \bibinfo{journal}{\emph{Data \& knowledge engineering}} \bibinfo{volume}{65}, \bibinfo{number}{1} (\bibinfo{year}{2008}), \bibinfo{pages}{126--146}.
\newblock


\bibitem[Thompson et~al\mbox{.}(2022)]%
        {thompson2022ancient}
\bibfield{author}{\bibinfo{person}{Amy~E Thompson}, \bibinfo{person}{John~P Walden}, \bibinfo{person}{Adrian~SZ Chase}, \bibinfo{person}{Scott~R Hutson}, \bibinfo{person}{Damien~B Marken}, \bibinfo{person}{Bernadette Cap}, \bibinfo{person}{Eric~C Fries}, \bibinfo{person}{M~Rodrigo Guzman~Piedrasanta}, \bibinfo{person}{Timothy~S Hare}, \bibinfo{person}{Sherman~W Horn~III}, {et~al\mbox{.}}} \bibinfo{year}{2022}\natexlab{}.
\newblock \showarticletitle{Ancient Lowland Maya neighborhoods: Average Nearest Neighbor analysis and kernel density models, environments, and urban scale}.
\newblock \bibinfo{journal}{\emph{PloS one}} \bibinfo{volume}{17}, \bibinfo{number}{11} (\bibinfo{year}{2022}), \bibinfo{pages}{e0275916}.
\newblock


\bibitem[Tian et~al\mbox{.}(2022)]%
        {tian2022survey}
\bibfield{author}{\bibinfo{person}{Ruijie Tian}, \bibinfo{person}{Huawei Zhai}, \bibinfo{person}{Weishi Zhang}, \bibinfo{person}{Fei Wang}, {and} \bibinfo{person}{Yao Guan}.} \bibinfo{year}{2022}\natexlab{}.
\newblock \showarticletitle{A survey of spatio-temporal big data indexing methods in distributed environment}.
\newblock \bibinfo{journal}{\emph{IEEE Journal of Selected Topics in Applied Earth Observations and Remote Sensing}}  \bibinfo{volume}{15} (\bibinfo{year}{2022}), \bibinfo{pages}{4132--4155}.
\newblock


\bibitem[Tritsarolis et~al\mbox{.}(2022)]%
        {tritsarolis2022piraeus}
\bibfield{author}{\bibinfo{person}{Andreas Tritsarolis}, \bibinfo{person}{Yannis Kontoulis}, {and} \bibinfo{person}{Yannis Theodoridis}.} \bibinfo{year}{2022}\natexlab{}.
\newblock \showarticletitle{The Piraeus AIS dataset for large-scale maritime data analytics}.
\newblock \bibinfo{journal}{\emph{Data in brief}}  \bibinfo{volume}{40} (\bibinfo{year}{2022}), \bibinfo{pages}{107782}.
\newblock


\bibitem[Wang and Shan(2005)]%
        {wang2005space}
\bibfield{author}{\bibinfo{person}{Jun Wang} {and} \bibinfo{person}{Jie Shan}.} \bibinfo{year}{2005}\natexlab{}.
\newblock \showarticletitle{Space filling curve based point clouds index}. In \bibinfo{booktitle}{\emph{Proceedings of the 8th International Conference on GeoComputation}}. \bibinfo{pages}{551--562}.
\newblock


\bibitem[Wang et~al\mbox{.}(2023)]%
        {wang2023slbrin}
\bibfield{author}{\bibinfo{person}{Lijun Wang}, \bibinfo{person}{Linshu Hu}, \bibinfo{person}{Chenhua Fu}, \bibinfo{person}{Yuhan Yu}, \bibinfo{person}{Peng Tang}, \bibinfo{person}{Feng Zhang}, {and} \bibinfo{person}{Renyi Liu}.} \bibinfo{year}{2023}\natexlab{}.
\newblock \showarticletitle{SLBRIN: a spatial learned index based on brin}.
\newblock \bibinfo{journal}{\emph{ISPRS International Journal of Geo-Information}} \bibinfo{volume}{12}, \bibinfo{number}{4} (\bibinfo{year}{2023}), \bibinfo{pages}{171}.
\newblock


\bibitem[Wang et~al\mbox{.}(2021)]%
        {wang2021survey}
\bibfield{author}{\bibinfo{person}{Sheng Wang}, \bibinfo{person}{Zhifeng Bao}, \bibinfo{person}{J~Shane Culpepper}, {and} \bibinfo{person}{Gao Cong}.} \bibinfo{year}{2021}\natexlab{}.
\newblock \showarticletitle{A survey on trajectory data management, analytics, and learning}.
\newblock \bibinfo{journal}{\emph{ACM Computing Surveys (CSUR)}} \bibinfo{volume}{54}, \bibinfo{number}{2} (\bibinfo{year}{2021}), \bibinfo{pages}{1--36}.
\newblock


\bibitem[Wu et~al\mbox{.}(2017)]%
        {wu2017apply}
\bibfield{author}{\bibinfo{person}{Tzuhsien Wu}, \bibinfo{person}{Jerry Chou}, \bibinfo{person}{Norbert Podhorszki}, \bibinfo{person}{Junmin Gu}, \bibinfo{person}{Yuan Tian}, \bibinfo{person}{Scott Klasky}, {and} \bibinfo{person}{Kesheng Wu}.} \bibinfo{year}{2017}\natexlab{}.
\newblock \showarticletitle{Apply block index technique to scientific data analysis and I/O systems}. In \bibinfo{booktitle}{\emph{2017 17th IEEE/ACM International Symposium on Cluster, Cloud and Grid Computing (CCGRID)}}. IEEE, \bibinfo{pages}{865--871}.
\newblock


\bibitem[Xu et~al\mbox{.}(2015)]%
        {xu2015gmobench}
\bibfield{author}{\bibinfo{person}{Jianqiu Xu}, \bibinfo{person}{Ralf~Hartmut G{\"u}ting}, {and} \bibinfo{person}{Xiaolin Qin}.} \bibinfo{year}{2015}\natexlab{}.
\newblock \showarticletitle{GMOBench: Benchmarking generic moving objects}.
\newblock \bibinfo{journal}{\emph{GeoInformatica}}  \bibinfo{volume}{19} (\bibinfo{year}{2015}), \bibinfo{pages}{227--276}.
\newblock


\bibitem[Zhang and El-Shaarawi(2010)]%
        {zhang2010spatial}
\bibfield{author}{\bibinfo{person}{Hao Zhang} {and} \bibinfo{person}{Abdel El-Shaarawi}.} \bibinfo{year}{2010}\natexlab{}.
\newblock \showarticletitle{On spatial skew-Gaussian processes and applications}.
\newblock \bibinfo{journal}{\emph{Environmetrics: The official journal of the International Environmetrics Society}} \bibinfo{volume}{21}, \bibinfo{number}{1} (\bibinfo{year}{2010}), \bibinfo{pages}{33--47}.
\newblock


\bibitem[Zhang and Ross(2020)]%
        {zhang2020exploiting}
\bibfield{author}{\bibinfo{person}{Wangda Zhang} {and} \bibinfo{person}{Kenneth~A Ross}.} \bibinfo{year}{2020}\natexlab{}.
\newblock \showarticletitle{Exploiting data skew for improved query performance}.
\newblock \bibinfo{journal}{\emph{IEEE Transactions on Knowledge and Data Engineering}} \bibinfo{volume}{34}, \bibinfo{number}{5} (\bibinfo{year}{2020}), \bibinfo{pages}{2176--2189}.
\newblock


\bibitem[Zheng(2015)]%
        {zheng2015trajectory}
\bibfield{author}{\bibinfo{person}{Yu Zheng}.} \bibinfo{year}{2015}\natexlab{}.
\newblock \showarticletitle{Trajectory data mining: an overview}.
\newblock \bibinfo{journal}{\emph{ACM Transactions on Intelligent Systems and Technology (TIST)}} \bibinfo{volume}{6}, \bibinfo{number}{3} (\bibinfo{year}{2015}), \bibinfo{pages}{1--41}.
\newblock


\bibitem[Zhu et~al\mbox{.}(2007)]%
        {zhu2007efficient}
\bibfield{author}{\bibinfo{person}{Qing Zhu}, \bibinfo{person}{Jun Gong}, {and} \bibinfo{person}{Yeting Zhang}.} \bibinfo{year}{2007}\natexlab{}.
\newblock \showarticletitle{An efficient 3D R-tree spatial index method for virtual geographic environments}.
\newblock \bibinfo{journal}{\emph{ISPRS Journal of Photogrammetry and Remote Sensing}} \bibinfo{volume}{62}, \bibinfo{number}{3} (\bibinfo{year}{2007}), \bibinfo{pages}{217--224}.
\newblock


\bibitem[Zim{\'a}nyi et~al\mbox{.}(2020)]%
        {zimanyi2020mobilitydb}
\bibfield{author}{\bibinfo{person}{Esteban Zim{\'a}nyi}, \bibinfo{person}{Mahmoud Sakr}, {and} \bibinfo{person}{Arthur Lesuisse}.} \bibinfo{year}{2020}\natexlab{}.
\newblock \showarticletitle{MobilityDB: A mobility database based on PostgreSQL and PostGIS}.
\newblock \bibinfo{journal}{\emph{ACM Transactions on Database Systems (TODS)}} \bibinfo{volume}{45}, \bibinfo{number}{4} (\bibinfo{year}{2020}), \bibinfo{pages}{1--42}.
\newblock


\end{thebibliography}

\end{document}